\documentstyle[prd,aps,preprint,epsfig]{revtex}

\newcommand{\beq}{\begin{eqnarray}}
\newcommand{\eeq}{\end{eqnarray}}

\begin{document}
\tighten

\preprint{
\noindent
\begin{minipage}[t]{3in}
\begin{flushright}
LBNL-41826\\
hep-ph/9805405 \\
\end{flushright}
\end{minipage}
}

\title{Radiative Decay of a Long-Lived Particle and Big-Bang
Nucleosynthesis}
\author{Erich Holtmann$^a$, Masahiro Kawasaki$^b$, Kazunori Kohri$^b$, and  
Takeo Moroi$^a$}

\address{$^a$Theoretical Physics Group,
     Lawrence Berkeley National Laboratory,\\
     University of California, Berkeley, California 94720}

\address{$^b$ Institute for Cosmic Ray Research, The University of Tokyo,
     Tanashi 188-8502, Japan}


\maketitle
\begin{abstract}
The effects of radiatively decaying, long-lived particles on
big-bang nucleosynthesis (BBN) are discussed. If high-energy
photons are emitted after BBN, they may change the abundances of
the light elements through photodissociation processes, which may
result in a significant discrepancy between the BBN theory and
observation.  We calculate the abundances of the light elements,
including the effects of photodissociation induced by a
radiatively decaying particle, but neglecting the hadronic
branching ratio.  Using these calculated abundances, we derive a
constraint on such particles by comparing our theoretical results
with observations.  Taking into account the recent controversies
regarding the observations of the light-element abundances, we
derive constraints for various combinations of the measurements.
We also discuss several models which predict such radiatively
decaying particles, and we derive constraints on such models.
\end{abstract}




\renewcommand{\thefootnote}{\arabic{footnote}}
\setcounter{footnote}{0}

\section{Introduction}
\label{sec:intro}

Big-bang nucleosynthesis (BBN) has been used to impose constraints 
on neutrinos and other hypothetical particles predicted by particle 
physics, because BBN is very sensitive to the thermal history of the 
early universe at temperatures $T \lesssim 1$ MeV~\cite{BBN-review}.

Weakly interacting, massive particles appear often in particle physics.
In this paper, we consider particles which have masses of $\sim
O(100{\rm ~GeV})$ and which interact with other particle only very
weakly ({\it e.g.}, through gravitation).  These particles have
lifetimes so long that they decay after the BBN of the light elements
(D, $^3$He, $^4$He, etc.), so they and their decay products may affect
the thermal history of the universe.  In particular, if the long-lived
particles decay into photons, then the emitted high-energy photons
induce electromagnetic cascades and produce many soft photons.  If
the energy of these photons exceeds the binding energies of the light
nuclides, then photodissociation may profoundly alter the light
element abundances.  Thus, we can impose constraints on the abundance
and lifetime of a long-lived particle species, by considering the
photodissociation processes induced by its decay.  There are many
works on this subject, such as the constraints on massive neutrinos
and gravitinos obtained by the comparison between the theoretical
predictions and
observations~\cite{Lindley,Raddec,KM1,KM2,hadron}.\footnote
 {As pointed out in Ref.~\cite{hadrBR}, even if the parent particle
decays only into photons, these photons will produce hadrons with a
branching ratio of at least 1\%.  However, since there is no data on
some crucial cross sections involving $^7$Li and $^7$Be, we cannot
include hadrodissociation in our statistical analysis.  Since we have
neglected hadrodissociation, our constraints may be regarded as
conservative bounds.}

A couple of years ago, Hata {\it et al.}~\cite{HSSTWBL} claimed that
light-element observations seemed
to conflict with the theoretical predictions of
standard BBN.  Their point was that standard BBN predicts too
much $^4$He, if the baryon number density is determined
by the D abundance inferred from observations;
equivalently, standard BBN
predicts too much D, if the baryon number density is
determined by the $^4$He observations.  Inspired by this ``crisis
in BBN,'' many people re-examined standard and non-standard BBN by
including systematic errors in the observations, or by introducing
some non-standard properties of neutrinos~\cite{xinu,mstau}.  In a
previous paper~\cite{PRLHKM}, we investigated the effect upon BBN
of radiatively decaying, massive particles. These particles induce
an electromagnetic cascade.  We found that in a certain
parameter region, the photons in this cascade destroy only D,
so that the predicted abundances of D, $^3$He, and $^4$He fit the
observations.

However, since the ``BBN crisis'' was claimed, the situation
concerning the observations of deuterium has changed.  The D
abundances in highly red-shifted quasar absorption systems (QAS) have
been observed. The abundance of D in high-$z$ QAS is considered to be
the primordial value. Thanks to these direct new observations, we no
longer need to use poorly understood models of chemical evolution to
infer the primordial abundance from the material in solar
neighborhood.

Moreover, there are also differing determinations of the primordial
$^4$He abundance.  Hata {\it et al.} used a relatively low $^4$He
abundance (viz., $Y\simeq 0.234$, where $Y$ is the primordial mass
fraction of $^4$He)~\cite{pagel,OliSkiSte}.  However, a higher $^4$He
abundance ($Y \simeq 0.244$) has also been
reported~\cite{IzoThuLip94,ThuIzo,IzoThuLip97}, and it has been noted
that this higher observation alleviates the discrepancy with standard
BBN theory~\cite{KerSar}.  The typical errors in $^4$He observations
are less than $\simeq 0.005$, so we have discordant data for $^4$He.

Since we have discordant $^4$He abundances and new observations for D,
the previous constraint on the radiative decay of long-lived particles
must be revised.  In addition, the statistical analyses on radiatively
decaying particles are insufficient in the previous works.  Therefore,
in our present paper, we perform a better statistical analysis of
long-lived, radiatively decaying particles, and of the resultant
photodissociations, in order to constrain the abundances and lifetimes
of long-lived particles.  In deriving the constraint, we use both high
and low values of the $^4$He abundance, because it is premature to
decide which data are correct. As a result, it will be shown that for
low values of the $^4$He abundance, we have a poor agreement
between the observations and the standard BBN theory. Moreover, we show in
this case that a long-lived particle with appropriate abundance and
lifetime can solve the discrepancy.  In the case of high
$^4$He, standard BBN
fits the observations, so we derive stringent constraints on the
properties of long-lived particles.

In this paper, we also include the photodissociations of $^7$Li and
$^6$Li for the first time. As we will show later, the destruction of
$^7$Li does not dramatically affect the predicted D and $^4$He, in the
region where the observed D and $^4$He values are best fit.  However,
the $^6$Li produced by the destruction of $^7$Li can be two orders of
magnitude more abundant than the standard BBN prediction of $^6$Li/H
$\sim O(10^{-12})$. We discuss the possibility that this process may
be the origin of the $^6$Li which is observed in some low-metallicity
halo stars.

In Sec.~\ref{sec:SBBN}, we study how consistent the theoretically
predicted abundances and observations are, in the case of standard
BBN.  The radiative decay of long-lived particles is considered in
Sec.~\ref{sec:BBNX}, and the particle physics models which predict such
long-lived particles are presented in Sec.~\ref{sec:model}.  Finally,
Sec.~\ref{sec:summary} is devoted to discussion and the conclusion.

\section{Standard Big-Bang Nucleosynthesis}
\label{sec:SBBN}

We begin by reviewing standard big-bang nucleosynthesis (SBBN).
We are interested in the light elements, since their
primordial abundances can be estimated from observations. In
particular, we check the consistency between the theoretical
predictions and the observations for the following quantities:
 \beq
  y_2 &=& n_{\rm D} / n_{{\rm H}},\\
  Y &=& \rho_{^4{\rm He}} / \rho_{{\rm B}}, \\
  y_6 &=& n_{^6{\rm Li}} / n_{{\rm H}}, \\
  r   &=& n_{^3{\rm He}} / n_{{\rm D}}, \\
  y_7 &=& n_{^7{\rm Li}} / n_{{\rm H}},
 \eeq
where  $\rho_{{\rm B}}$ is the total baryon energy density.

In this section, we first review the observations of the light
elements, and the extrapolations back to the primordial abundances.
Next, we describe our theoretical calculations of these abundances, by
using standard big-bang theory as an example.  Finally, we compare the
theoretical and observed light-element abundances to determine how
well the SBBN theory works.

\subsection{Review of Observation}
\label{sect-obs}

Let us start with a review of the observations of the light-element
abundances. Two factors complicate the interpretation of the
observations of the light-element abundances.  First, there are
various observational determinations
for $^4$He which are not consistent with each
other, within the quoted errors.  This fact suggests that some groups
have underestimated their systematic error.\footnote{
 It is also possible that primordial nucleosynthesis
was truly inhomogeneous~\cite{inhomo}.
 However, in this paper we adopt the conventional belief that BBN was
 homogeneous.
} 
 We believe it is
premature to judge which measurements are reliable; hence, we consider
both of the observations when we test the
consistency between theory and observation.  Second, some guesswork is
involved in the extrapolation back from the observed values to the
primordial values, as we shall discuss below. Keeping these factors in
mind, we review the estimations of the primordial abundances of D,
$^3$He, $^4$He, $^6$Li, and $^7$Li.

D/H has been measured in the absorption lines of highly red-shifted
(and therefore presumably primordial) H$_{\rm I}$ (neutral hydrogen)
clouds which are backlit by quasars. The latest result
suggests~\cite{BurTyt}
 \begin{eqnarray}
   y^{obs}_2 = (3.39 \pm 0.25) \times 10^{-5}.
 \label{lowD}
 \end{eqnarray}
We use this value rather than the higher
abundance which had been
reported~\cite{songaila,earlyQASH,no-interloper,higp}
in some of the old measurements of D/H.  The older
results suggested the abundance was $y_2\sim O(10^{-4})$. However,
we believe these results to be much more uncertain.
For example, the authors
of Ref.~\cite{no-interloper} admitted a large uncertainty in their
results. Furthermore, results given in Ref.~\cite{higp} are based on the fit
of only the Lyman alpha limit, and the resolution is not good. Therefore,
we will not use the high D values in deriving the constraints, but we
will just discuss the implications of taking the high value
seriously.

For $^3$He, we use the pre-solar measurements.  In this paper, we
do not rely upon any models of galactic and stellar chemical
evolution, because of the large uncertainty involved in extrapolating
back to the primordial abundance. But it is reasonable to
assume that $^3$He/D is an increasing function of
time, because D is the most fragile isotope, and it is certainly destroyed
whenever $^3$He is destroyed. Using the solar-system data reanalyzed
by Geiss~\cite{Geiss93}, the $^3$He/D ratio is estimated
to be~\cite{Sigl}
\begin{equation}
    \label{geiss}
    r^{obs}_\odot \equiv
    \left(y^{obs}_3/y^{obs}_2\right)_{\odot} = 0.591 \pm 0.536,
 \label{He3/D}
\end{equation}
where $\odot$ denotes the pre-solar abundance.
We take this to be an upper bound on the primordial $^3$He to D ratio:
\begin{equation}
    \label{upper_bound32}
    r^{obs} < r^{obs}_{\odot}.
\end{equation}
Because the theoretical prediction of $^3$He/D in SBBN agrees
so well with this upper bound, we do not include this
constraint in the SBBN analysis. But when we
investigate the photodissociation scenario, the situation is quite
different. $^4$He photodissociation produces both D and $^3$He and
can raise the $^3$He to D ratio~\cite{Sigl}.
Hence, in our analysis of BBN
with photodissociation, we include this upper bound, as described
in the appendix (see Eq. (\ref{eqn:p23})).
An analysis based upon the chemical evolution of $^3$He and D
will appear in a separate paper by one of the authors~\cite{holtmann}.

The primordial $^4$He abundance is deduced from observations of
extragalactic H$_{\rm II}$ regions (clouds of ionized hydrogen).
Currently, there are two classes of $Y^{obs}$, reported by several
independent groups of observers.  Hence, we consider two cases: one
low, and one high.

We take our low $^4$He abundance from Olive, Skillman,
and Steigman~\cite{OliSkiSte}.
They used measurements of $^4$He and O/H
in 62 extragalactic H$_{\rm II}$ regions, and linearly extrapolated
back to O/H$= 0$ to deduce the primordial value
\beq
\mbox{\rm Low: }
Y^{obs} = 0.234 \pm (0.002)_{stat} \pm (0.005)_{syst}   \label{lowHe}.
\eeq
(When they restrict their data set to only the lowest metallicity
data, they obtain $Y^{obs}= 0.230 \pm 0.003$.)  Their
systematic error comes from numerous sources, but they claim that no
source is expected to be much more than 2\%.  In particular, they
estimate that stellar absorption is of order 1\% or less.

We take our high $^4$He abundance from Thuan and
Izotov~\cite{ThuIzo}.
They used measurements of $^4$He and O/H in a new sample of 45 blue
compact dwarf galaxies to obtain
\beq
\mbox{\rm High: }
Y^{obs} = 0.244 \pm (0.002)_{stat} \pm (0.005)_{syst}   \label{highHe}.
\eeq
The last error is an estimate of the systematic error, taken from
Izotov, Thuan, and Lipovetsky~\cite{IzoThuLip97}.
Thuan and Izotov claim that He$_{\rm I}$ stellar absorption is an
important effect; this explains some of the difference between their
result and that of Olive, Skillman, and Steigman.

Rather than attempting to judge which group has done a better job of
choosing their sample and correcting for systematic errors, we
prefer to remain open-minded.  Hence, we shall use both the high and
low $^4$He abundances, without expressing a preference for one over
the other.

The $^7$Li/H abundance is taken from observations of the surfaces of
Pop II (old generation) halo stars.
$^7$Li is a fragile isotope and is easily destroyed in the warmer interior
layers of a star.  Since more massive (or equivalently, hotter) stars are
mixed less, one might hope that the surfaces of old, hot stars consist of
primordial material.  Indeed, Spite and Spite~\cite{SpiSpi} discovered
a ``plateau'' in the graph of $^7$Li abundance {\it vs.} temperature of old halo
stars, at high temperature.  This plateau is interpreted as evidence that
truly primordial $^7$Li has been detected.
Using data from 41
plateau stars, Bonifacio and Molaro~\cite{BonMol} determine the
primordial value $\log_{10}(y^{obs}_7) = -9.762 \pm (0.012)_{stat} \pm
(0.05)_{syst}$.  Bonifacio and Molaro argue that the data provides no
evidence for $^7$Li/H depletion in the stellar atmospheres (caused by,
{\it e.g.}, stellar winds, rotational mixing, or diffusion).  However,
for our analysis, we shall adopt the more cautious estimate of
Hogan~\cite{Hog} that $^7$Li may have been supplemented (by production
in cosmic-ray interactions) or depleted (in stars) by a factor of
two:~\cite{factor-of-two}
 \beq
  \log_{10}(y^{obs}_7) = 
  -9.76 \pm (0.012)_{stat} 
  \pm (0.05)_{syst}
  \pm (0.3)_{factor~of~2}.
 \label{Li7}
 \eeq

Because $^6$Li is so much rarer than $^7$Li, it is much more difficult
to observe.  Currently, there is insufficient data to find the ``Spite
plateau'' of $^6$Li.  However, we can set an upper bound on
$^6$Li/$^7$Li, since it is generally agreed that the evolution of
$^6$Li is dominated by production by spallation (reactions
of cosmic rays with the interstellar medium). The upper bounds
on $^6$Li/$^7$Li observed in low-metallicity
([Fe/H] $\leq -2.0$) halo stars range from~\cite{HobTho}
$y_6/y_7 \lesssim 0.045$ to $y_6/y_7 \lesssim 0.13$. (Note that
the primordial $^6$Li and $^7$Li have both been destroyed in material
which has been processed by stars.)

Rotational mixing models~\cite{SmiLamNis} yield a survival
factor for $^7$Li of order 0.05 and a survival factor for $^6$Li
of order 0.005.  Therefore, the upper bound for primordial
$^6$Li/$^7$Li ranges approximately from
 \beq
  y^{obs}_6/y^{obs}_7 \lesssim 0.5 \ \ {\rm to} \ \ 1.3.
 \label{Li6}
 \eeq
Since we have only a rough range of upper bounds on $^6$Li, and no
lower bound, we will not use $^6$Li in our statistical analysis
to test the concordance between observation and theory.
Instead, we will just check the consistency of our theoretical
results with the above constraint.

\subsection{Theoretical Calculations}

Theoretically, the primordial abundances of the light elements
in SBBN
depend only upon a single parameter: the baryon-to-photon ratio
$\eta$. In our analysis, we modified Kawano's nucleosynthesis
code~\cite{Kaw} to calculate the primordial light-element
abundances and uncertainties.

In our calculation, we included the uncertainty in the neutron
lifetime~\cite{PDG}, in the rates of the 11 most important
nucleosynthesis reactions~\cite{SmiKawMal}, and in the rates of
the photofission reactions (see Table~\ref{table:pf}).
We treated the neutron lifetime, the nucleosynthesis reaction rates,
and the photofission reaction rates
as independent random variables with Gaussian
probability density functions (p.d.f.'s).  We performed a
Monte-Carlo\footnote{
It has recently been demonstrated that the uncertainties in SBBN
can be quantified by the much quicker method of linear propagation
of errors.~\cite{linprop}}
over the neutron lifetime, nucleosynthesis reaction rates,
and photofission reaction rates, and we
found that the light-element abundances were distributed approximately
according to independent, Gaussian p.d.f.'s. Therefore, the
p.d.f. $p_{tot}^{th}$ for the theoretical abundances is given by the
product of the Gaussian p.d.f.'s
\beq
p^{Gauss}(x;\bar{x},\sigma) &=&
 \frac{1}{\sqrt{2 \pi}\sigma}
 \exp \left[ -\frac{1}{2}
            \left( \frac{ x-\bar{x} }{\sigma} \right)^2
      \right]
\eeq
for the individual
abundances:
\beq
p_{tot}^{th} (y^{th}_2,Y^{th}, \log_{10}y^{th}_7)
 = p^{Gauss}_2(         y^{th}_2) \times
   p^{Gauss}_4(         Y^{th}  ) \times
   p^{Gauss}_7(\log_{10}y^{th}_7).
\eeq

In Fig.~1,  we have plotted the theoretical predictions for the
light-element abundances (solid lines) with their one-sigma errors
(dashed lines), as functions of $\eta$.

The dependences of the abundances on $\eta$ can be seen
intuitively~\cite{BBN-review,SBBN}. The $^4$He abundance is a gentle,
monotonically increasing function of $\eta$. As $\eta$ increases,
$^4$He is produced earlier because the ``deuterium bottleneck'' is
overcome at a higher temperature due to the higher baryon
density. Fewer neutrons have had time to decay, so more $^4$He is
synthesized.  Since $^4$He is the most tightly bound of the light
nuclei, D and $^3$He are fused into $^4$He. The surviving abundances
of D and $^3$He are determined by the competition between their
destruction rates and the expansion rate.  The destruction rates are
proportional to $\eta$, so the larger $\eta$ is, the longer the
destruction reactions continue.  Therefore, D and $^3$He are
monotonically decreasing functions of $\eta$. Moreover, the slope of D
is steeper, because the binding energy of D is smaller than $^3$He.
 
The graph of $^7$Li has a ``trough'' near $\eta \sim 3 \times
10^{-10}$.  For a low baryon density $\eta \lesssim 3 \times
10^{-10}$, $^7$Li is produced by $^4$He(T, $\gamma$)$^7$Li and is
destroyed by $^7$Li(p, $\alpha$)$^4$He. As $\eta$ increases, the
destruction reaction become more efficient and the produced $^7$Li
tends to decrease. On the other hand for a high baryon density $\eta
\gtrsim 3 \times 10^{-10}$, $^7$Li is mainly produced through the
electron capture of $^7$Be, which is produced by $^3$He($\alpha$,
$\gamma$)$^7$Be.  Because $^7$Be
production becomes more effective as $\eta$ increases, the synthesized
$^7$Li increases. The ``trough'' results from the overlap of these two
components. The dominant source of $^6$Li in SBBN is D($\alpha$,
$\gamma$)$^6$Li. Thus, the $\eta$ dependence of $^6$Li resembles that
of D.

We have also plotted the 1-$\sigma$ observational constraints. The
amount of overlap of the boxes is a rough measure of the consistency
between theory and observations.  We can also see the favored ranges
of $\eta$. However, we will discuss the details of our analysis more
carefully in the following section.

\subsection{Statistical Analysis and Results}
\label{sec-bbnResults}

Next, let us briefly explain how we quantify the consistency
between theory and observation. For this purpose, we define the
variable $\chi^2$ as
\beq
 \chi^2 = \sum_i \frac{(a_i^{th} - a_i^{obs})^2}
                      {(\sigma_i^{th})^2 + (\sigma_i^{obs})^2} \label{chi2}
\eeq
where $a_i = (y_2, Y, \log_{10}y_7)$,
and we add the systematic errors in quadrature:
$(\sigma_i^{obs})^2=(\sigma_i^{syst})^2+(\sigma_i^{stat})^2$.  (See
the appendix for a detailed explanation of our use of
$\chi^2$.)  $\chi^2$ depends upon the parameters of our theory
(viz. $\eta$ in SBBN) through $a_i^{th}$ and $\sigma_i^{th}$.

Notice that we do not include $^6$Li in the calculation of $\chi^2$,
since the $^6$Li abundance has not been measured well. Instead, we
check that $y^{th}_6/y^{th}_7$ satisfies the bound (\ref{Li6}).  In
the case of SBBN, we found that the $^6$Li abundance is small enough
over the entire range of $\eta$ from $8.0 \times 10^{-11}$ to $1.0
\times 10^{-9}$.  (Numerically, $y^{th}_6/y^{th}_7 < 5\times 10^{-4}$,
which is well below the bound (\ref{Li6}).)

With this $\chi^2$ variable, we discuss how well the theoretical
prediction agrees with observation. More precisely, we calculate from
$\chi^2$ the confidence level (C.L.)  with which the SBBN theory is
excluded, at a given point in the parameter space of our theory
(for three degrees of freedom):
 \beq
  {\rm C.L.} &=& \int_0^{\chi^2}
  \frac{1}{2^{3/2}\Gamma(\frac32)}y^{\frac12}e^{- \frac{y}{2}} d y
  \label{chidis}\\
&=&
  -\sqrt{\frac{2 \chi^2}{\pi}}
  \exp \left( -\frac{\chi^2}{2} \right)
  + {\rm erf} \left( \sqrt{\frac{\chi^2}{2}} \right),
 \label{CL}
 \eeq

In Fig.~2, we have plotted the $\chi^2$ and confidence level at which
SBBN theory is excluded by the observations, as a function of $\eta$.
We find that high $^4$He is allowed at better than the 68\%
C.L. at $\eta \sim 5 \times 10^{-10}$, while for low $^4$He, no value
of $\eta$ works at the 91.5\% C.L.

The case of low $^4$He suggests a discrepancy with standard BBN.  Some
people believe that this casts doubt on the low D or low $^4$He
measurements\cite{CopSchTur}.  However, we do not want to assume SBBN
theory and use it to judge the validity of the observations; rather,
we use the observations to test BBN theory.  Therefore, we give equal
consideration to both the high and low observed abundances
of $^4$He.

Before closing this section, we apply our analysis to constrain the
number of neutrino species. Here, we vary $\eta$ and the number
$N_\nu$ of neutrino species, and we calculate the confidence level as
a function of these variables. The results are shown in
Fig.~3a,b. We can see that the standard scenario ($N_\nu=3$)
results in a good fit with a high value of $^4$He, while the case of low
$^4$He results in a discrepancy. In fact, low $^4$He prefers $N_\nu
\sim 2$, as pointed out by \cite{HSSTWBL,mstau}. In
Table~\ref{table:nueta}, we show the 95$\%$ C.L. bounds for the number
of neutrino species $N_\nu$ and $\eta$ in the two cases of the $^4$He
abundances.


\section{BBN + $X$}
\label{sec:BBNX}

In this section, we discuss the implications of a radiatively
decaying particle $X$ for BBN. For this purpose, we first
discuss the behavior of the photon spectrum induced by $X$. Then we
show the abundances of the light elements, including the effects of the
photodissociation induced by $X$. Comparing these abundances with
observations, we constrain the parameter space for $\eta$ and $X$.

\subsection{Photon Spectrum}

In order to discuss the effect of high-energy photons on BBN, 
we need to know the shape of the photon spectrum induced by the
primary high-energy photons from $X$ decay.

In the background thermal bath (which, in our case, is a mixture of
photons $\gamma_{\rm BG}$,
electrons $e^-_{\rm BG}$, and
nucleons $N_{\rm BG}$), high-energy photons lose their
energy by various cascade processes. In the cascade, the photon
spectrum is induced, as discussed in various literature\cite{spectrum}
. The important
processes in our case are:
 \begin{itemize}
  \item Double-photon pair creation
($\gamma +\gamma_{\rm BG} \rightarrow e^+ +e^-$)
  \item Photon-photon scattering
($\gamma +\gamma_{\rm BG} \rightarrow \gamma +\gamma$)
  \item Pair creation in nuclei
($\gamma  +N_{\rm BG} \rightarrow e^+  +e^- + N$)
  \item Compton scattering
($\gamma  +e^-_{\rm BG} \rightarrow \gamma  +e^-$)
  \item Inverse Compton scattering
($e^{\pm} +\gamma_{\rm BG} \rightarrow e^{\pm}  +\gamma$)
 \end{itemize}
(We may neglect double Compton scattering
$ \gamma + e^-_{\rm BG} \rightleftharpoons \gamma + \gamma + e^- $,
because Compton scattering is more important for thermalizing
high-energy photons.)
In our analysis, we numerically solved the Boltzmann equation
including the above effects, and obtained the distribution function of
photons, $f_\gamma (E_\gamma)$. (For details, see
Refs.~\cite{KM1,KM2}.)

In Fig.~4, we show the photon spectrum for several
temperatures $T$. Roughly speaking, we can see a large drop-off at
$E_\gamma\sim m_e^2/22T$ for each temperature.  Above this threshold,
the photon spectrum is extremely suppressed.

The qualitative behavior of the photon spectrum can be understood in
the following way. If the photon energy is high enough, then
double-photon pair creation is so efficient that this process
dominates the cascade. However, once the photon energy becomes much
smaller than $O(m_e^2/T)$, this process is kinematically
blocked. Numerically, this threshold is about $m_e^2/22T$, as we
mentioned. Then, photon-photon scattering dominates.  However, since
the scattering rate due to this process is proportional to
$E_\gamma^3$, photon-photon scattering becomes unimportant in the
limit $E_\gamma\rightarrow 0$.  Therefore, for $E_\gamma\ll
O(m_e^2/T)$, the remaining processes (pair creation in nuclei and
inverse Compton scattering) are the most important.

The crucial point is that the scattering rate for $E_\gamma\gtrsim
m_e^2/22T$ is much larger than that for $E_\gamma\ll m_e^2/22T$, since
the number of targets in the former case is several orders of
magnitude larger than in the latter. This is why the photon spectrum
is extremely suppressed for $E_\gamma\gtrsim m_e^2/22T$. As a result,
if the $X$ particle decays in a thermal bath with temperature
$T\gtrsim m_e^2/22Q$ (where $Q$ is the binding energy of a nuclide)
then photodissociation is not effective.
 
\subsection{Abundance of Light Elements with $X$}

Once the photon spectrum is formed, it induces the photodissociation
of the light nuclei, which modifies the result of SBBN. This process is
governed by the following Boltzmann equation:
 \begin{eqnarray}
  \frac{dn_N}{dt} + 3Hn_N &=& \left[\frac{dn_N}{dt}\right]_{\rm SBBN}
  - n_N \sum_{N'}\int dE_\gamma 
  \sigma_{N\gamma\rightarrow N'}(E_\gamma) f_\gamma (E_\gamma)
 \nonumber \\ &&
  + \sum_{N''}n_{N''} \int dE_\gamma 
  \sigma_{N''\gamma\rightarrow N} (E_\gamma) f_\gamma (E_\gamma),
 \end{eqnarray}
 where $n_N$ is the number density of the nuclei $N$, and
$[dn_N/dt]_{\rm SBBN}$ denotes the SBBN contribution to the Boltzmann
equation. To take account of the photodissociation processes, we
modified the Kawano code~\cite{Kaw}, and calculated the abundances of
the light elements. The photodissociation processes we included in our
calculation are listed in Table~\ref{table:pf}.

The abundances of light nuclides will be functions of the lifetime of
$X$ ($\tau_X$), the mass of $X$ ($m_X$), the abundance of $X$ before
electron-positron annihilation 
\begin{equation}
    \label{yx}
 Y_X = n_X/n_\gamma,
\end{equation}
and the baryon-to-photon ratio ($\eta$).  In our numerical BBN
simulations, we found that the nuclide abundances depend only on the
mass abundance $m_X Y_X$, not on $m_X$ and $Y_X$ separately.  In
Figs.~5 -- 9, we show the abundances
of light nuclei in the $m_X Y_X$ vs. $\tau_X$ plane, at fixed $\eta$.

We can understand the
qualitative behaviors of the abundances in the following way.
First of all, if the mass density
of $X$ is small enough, then the effects of $X$ are negligible, and hence we
reproduce the result of SBBN. Once the mass density gets larger, the SBBN
results are modified. The effects of $X$ strongly depend on $\tau_X$,
the lifetime of $X$. As we mentioned in the previous section, photons with
energy greater than $\sim m_e^2/22T$ participate in pair creation
before they can induce photofission. Therefore, if the
above threshold energy is smaller than the nuclear binding energy,
then photodissociation is not effective.

If $\tau_X\lesssim 10^4$~sec, then $m_e^2/22T \lesssim 2$MeV at the decay time of $X$, and photodissociation is negligible for all elements.
In this case, the main effect of $X$ is on the $^4$He
abundances: if the abundance of $X$ is large, its energy density
speeds up the expansion rate of the universe, so the neutron
freeze-out temperature becomes higher. As a result, $^4$He abundance is
enhanced relative to SBBN.

If $10^4$~sec $\lesssim\tau_X\lesssim 10^6$~sec, then 2~MeV $\lesssim
m_e^2/22T\lesssim$ 20~MeV.  In this case, $^4$He remains intact,
but D is effectively photodissociated through the process
${\rm D}+\gamma\rightarrow p+n$.  When $\tau_X \gtrsim 10^5$~sec,
$m_e^2/22T \gtrsim$ 7.7~MeV (the binding energy of $^3$He), so
$^3$He is dissociated for $\tau_X \sim 10^5$~sec and large enough
abundances $m_X Y_X \gtrsim 10^{-8}$~GeV.  However, D is even more fragile
than $^3$He, so the ratio $^3$He/D actually {\it increases} relative to SBBN
in this region,
since it is dominated by D destruction.
If the lifetime is long enough ($\tau_X\gtrsim 10^6$~sec), $^4$He can
also be destroyed effectively. In this case, the destruction of even a
small fraction of the $^4$He can result in significant production of
D and $^3$He, since the $^4$He abundance is
originally much larger than that of
D. This can be seen in Figs.~5 and 6:
for $\tau_X\gtrsim 10^6$~sec and $10^{-10}$~GeV $\lesssim
m_XY_X\lesssim 10^{-9}$~GeV, the abundance of D changes drastically
due to the photodissociation of $^4$He.  Moreover, two-body decays of
$^4$He into $^3$He or T (which decays into $^3$He)
are preferred over the three-body decay
$^4{\rm He} + \gamma \rightarrow p + n + {\rm D}$,
so the $^3$He/D ratio increases, relative to SBBN.
If $m_XY_X$ is large enough,
all the light elements are destroyed efficiently, resulting in very
small abundances.

So far, we have discussed the theoretical calculation of the light-element
abundances in a model with $X$ decay. In the next section, we compare the
theoretical calculations with observations, and derive constraints on
the properties of $X$.

\subsection{Comparison with Observation}
\label{sec:compar}

Now, we compare the theoretical calculations with the observed
abundances and show how we can constrain the model parameters. As we
mentioned in Section~\ref{sect-obs}, we have two $^4$He values which
are inferred from various observed data to be the primordial
components. We will consider both cases and derive a constraint.
(The statistical analysis we use to calculate the confidence level
is explained in the appendix.)

Let us start our discussion with the low $^4$He case
($Y^{obs}=0.234\pm (0.002)_{stat}\pm (0.005)_{syst}$). Recalling
that the low observed $^4$He value did not result in a good fit in the
case of SBBN, we search the parameter space for regions of better fit
than we can obtain with SBBN.

In Fig.~10, we show the contours of the confidence level computed using
four abundances
(D/H, $^4$He, $^3$He/D, and $^7$Li/H),
for some representative $\eta$
values ($\eta_{10}=2,4,5,6$), where
 \begin{eqnarray}
  \eta_{10}\equiv \eta \times 10^{10}.
 \end{eqnarray}

The region of parameter space which is allowed at the 68\%
C.L. extends down to low $\eta$ (see Fig.10a).  Near $\eta_{10}=2$,
deuterium is destroyed by an order of magnitude (without net
destruction of $^4$He), so that the remaining deuterium agrees with
the calculated low $^4$He.  For $\eta_{10} \sim 5$,
SBBN ($m_X Y_X = 0$) is allowed, and we have an approximate
upper bound on $m_X Y_X$ (although for $\tau_X < 10^6$ sec, slightly
higher values of $m_X Y_X$ are allowed at low $\eta$).
For $\eta_{10} \gtrsim 6$,
no region is allowed at the 95\% C.L., because $\eta$ becomes
to high to match even the observed D.
We also plotted the regions excluded by the
observational upper bounds on $^6$Li/$^7$Li. The shaded regions are
$y_6/y_7 \gtrsim 0.5$, and the darker shaded regions are $y_6/y_7
\gtrsim 1.3$. Even if we adopt the stronger bound $y_6/y_7 \lesssim
0.5$, our theoretical results are consistent with the observed $^6$Li
value.

In Fig.~11, we show the contours of the confidence levels for various
lifetimes, $\tau_X = 10^4, 10^5, 10^6$ sec. As the lifetime decreases,
the background temperature at the time of decay increases, so the
threshold energy of double-photon pair creation decreases. Then for a
fixed $m_X Y_X$, the number of photons contributing to D destruction
decreases. Thus, for shorter lifetimes, we need larger $m_X Y_X$ in
order to destroy sufficient amounts of D. The observed abundances
prefer non-vanishing $m_X Y_X$.

In Fig.~12, we show contours of $\chi^2$ which have been projected
along the $\eta$ axis into the $\tau_X$ - $m_X Y_X$ plane.  By
projection, we mean taking the lowest C.L. value along the $\eta$ axis
for a fixed point ($\tau_X$, $m_X Y_X$).
The region above the solid like is excluded at the
95\% C.L., while only the region within
the dotted line is allowed at the 68\% C.L.
The 95\% C.L. constraint for
$\tau_X \lesssim 10^6$~sec comes primarily from destruction of
too much D; for $\tau_X \gtrsim 10^6$~sec, it comes primarily from
overproduction of $^3$He in $^4$He photofission.

The lower $m_X Y_X$ region, {\it i.e.} $m_X Y_X\sim 10^{-14}$ GeV,
corresponds to SBBN, since there are not enough photons to affect the
light-element abundances. It is notable that these regions are outside
of the 68\% C.L. This fact may suggest the existence of a long-lived
massive particle $X$ and may be regarded as a hint of physics beyond
the standard model or standard big bang cosmology.

For example, in Fig.~13 we show the predicted abundances of $^4$He, D,
$^7$Li and $^6$Li adopting the model parameters $\tau_X = 10^6$ sec
and $m_X Y_X = 5 \times 10^{-10}$ GeV. The predicted abundances of
$^4$He and $^7$Li are nearly the same as in SBBN.  Only D is
destroyed; its abundance decreases by about 80\%.  At low $\eta \sim
(1.7-2.3) \times 10^{-10}$ in this model, the predicted abundances of
these three elements agree with the observed values.  It is
interesting that the produced $^6$Li abundance can be two orders of
magnitude larger than the SBBN prediction in this parameter
region. The origin of the observed $^6$Li abundance, $^6$Li/H~$\sim
O(10^{-12})$ is usually explained by cosmic ray spallation; however,
our model demonstrates the possibility that $^6$Li may have been
produced by the photodissociation of $^7$Li at an early epoch. Our
$^6$Li prediction is consistent with the upper bound Eq.~(\ref{Li6}).

Although $m_X Y_X \gtrsim 10^{-10}$~GeV is favored,
it is worth noting that SBBN lies within the 95\% C.L.
agreement between theory and observation.  In Fig.~12, the 95\% bound
for $\tau_X \lesssim 10^6$ sec comes from the constraint that not much
more than 90\% of the deuterium should be destroyed; for $\tau_X
\gtrsim 10^6$ sec the constraint is that deuterium should not be
produced from $^4$He photofission.  In Table~\ref{table:ll}, we show
the representative values of $m_X Y_X$ which correspond to 68\% and
95\% confidence levels respectively, for $\tau_X = 10^4 - 10^9$ sec.

Next, we would like to discuss the high $^4$He case ($Y^{obs} =
0.244 \pm (0.002)_{stat} \pm (0.005)_{syst}$). Since the D abundance
(\ref{lowD}) and high value of $^4$He (\ref{highHe})
both prefer a relatively high
value of $\eta$, the SBBN prediction can be consistent with observation in
this case. Therefore, we expect to be able to constrain the model
parameters.

For four representative $\eta$ values ($\eta_{10}=2, 4, 5, 6$), we
plot the contours of the confidence level in Fig.~14. In Fig.~2, we
see that the SBBN calculations agree with the observed abundances for
mid-range values of the baryon-to-photon ratio ($\eta \sim 5 \times
10^{-10}$).  Thus, the approximate
upper bound for $m_X Y_X$ is plotted in
Fig.~14c.  (Again, for $\tau_X < 10^6$ sec, slightly
higher values of $m_X Y_X$ are allowed at low $\eta$.)
In Fig.~15, we show the C.L. plots
for three typical lifetimes, $\tau_X = 10^4, 10^5, 10^6$ sec.
This plot shows that SBBN works at better than the 68\% C.L.
for a range of lifetimes, but the non-standard scenarios with
large $m_X Y_X$ and small $\eta$ do not work as well
as they did in the low $Y^{obs}$ case.
Finally, we show the C.L.  contours projected along the $\eta$ axis
into the $\tau_X$ - $m_X Y_X$ plane (Fig.~16).  Table~\ref{table:lh}
gives the upper bounds on $m_X Y_X$ (GeV) which correspond to 68\% and
95\% C.L., for some typical values of the lifetime.

Before we discuss additional constraints,
let us comment on the case of high values of D/H,
suggested by old observations~\cite{songaila,earlyQASH,no-interloper,higp}.
Even though the high values
seem less reliable, we believe their possibility has not been
completely ruled out. Therefore, it may be useful to comment on this
case.  The high value of D abundance [$y_2\sim O(10^{-4})$] prefers a
low value of $\eta$, and hence it is completely consistent with the
low value of $Y^{obs}$ in SBBN. Furthermore, if we adopt the relatively large
error bar for $y_2$ suggested by the observation, SBBN may also be
consistent with high $Y^{obs}$. Then, in this case also, we can
obtain upper bounds on the mass density $m_XY_X$ as a function of its
lifetime. The upper bound behaves like the case of low D and low
$^4$He shown in Fig.~16, and the upper bounds are
within a factor of ten for most of lifetimes.

\subsection{Additional Constraints}

We now mention additional constraints on our model.
Since the
cosmic microwave background radiation (CMBR) has been observed by
COBE~\cite{fixsen} to very closely follow a blackbody spectrum,
one may be concerned about the
constraint this gives on particles with lifetime longer than
$\sim 10^6$ sec~\cite{PRL70-2661}, which is when the double Compton
process ($ \gamma + e^- \rightleftharpoons \gamma + \gamma + e^- $)
freezes out~\cite{lightman}.\footnote
 {This constraint applies only to particles with lifetime shorter
than $\sim 4 \times 10^{10}$ sec, which corresponds to the decoupling
time of Compton/inverse Compton scattering.  After this time, injected
photons do not thermalize with the CMBR.}
 After this time, photon number
is conserved, so photon injection from a radiatively decaying
particle would cause the spectrum of the CMBR to become a
Bose-Einstein distribution with a finite chemical potential $\mu$.
COBE~\cite{fixsen} observations give us the constraint
$|\mu| \lesssim 9.0 \times 10^{-5}$. For small $\mu$,
the ratio of the injected to total photon energy density is
given by $\delta \rho_{\gamma}/\rho_{\gamma} \sim 0.71 \mu$.
Thus, we have the constraint
\begin{equation}
    \label{COBE}
    m_X Y_X \lesssim 6 \times 10^{-10} {\rm GeV}
          \left(\frac{\tau_X}{10^6{\rm sec}}\right)^{-\frac12} \quad
          {\rm for} \ 
     10^6 {\rm sec} \lesssim \tau_X \lesssim 4 \times 10^{10}
     {\rm sec}.  
\end{equation}
Note that the CMBR constraint is not as strong as the bounds 
we have obtained from BBN.  In particular, $^3$He/D gives us
our strongest constraint for lifetimes longer than
$10^6$ sec, because $^4$He photofission overproduces $^3$He~\cite{Sigl}.

In this paper, we have considered only radiative decays; {\it i.e.},
decays to photons and invisible particles.  If $X$ decayed to charged
leptons, the effects would be similar to those of the
decay to photons, because
charged leptons also generate electromagnetic cascades,
resulting in many soft photons. On
the other hand, if $X$ decayed only to neutrinos, the constraints
would become much weaker. In the minimal supersymmetric standard model (MSSM),
the $X$ particle would decay
to neutrinos and sneutrinos. The emitted neutrinos would scatter off of
background neutrinos, producing electron-positron pairs, which
would trigger an
electromagnetic cascade. Because the interaction between the emitted
and background neutrinos is weak, the destruction of the
light elements does not occur very efficiently~\cite{chglep}.
In contrast, if $X$ decayed to hadrons, we expect that our bounds would
tighten, because hadronic showers could be a significant source of D,
$^3$He, $^6$Li, $^7$Li, and $^7$Be~\cite{hadron}.  In fact, even
though we have assumed that $X$ decays only to photons, these photons
may convert to hadrons.  Thus, the branching ratio to hadrons is at
least of order 1~\%, if kinematically allowed~\cite{hadrBR}.  Since we
have neglected this effect,
our photodissociation bounds are conservative.

\section{Models}
\label{sec:model}

So far, we have discussed general constraints from BBN on radiatively
decaying particles. In the minimal standard model, there is no
such particle. However, some extensions of the standard model
naturally result in such exotic particles, and the light-element
abundances may be
significantly affected in these cases. In this section, we present
several examples of such radiatively decaying particles, and discuss
the constraints.

Our first example is the gravitino $\psi$, which appears in all the
supergravity models. The gravitino is the superpartner of the graviton,
and its interactions are suppressed by inverse powers of the reduced
Planck scale $M_*\simeq 2.4\times 10^{18}$ GeV.  Because of this
suppression, the lifetime of the gravitino is very long. Assuming that
the gravitino's dominant decay mode is to a photon and its
superpartner (the photino), the gravitino's
lifetime is given by
 \begin{eqnarray}
  \tau_{3/2}\simeq 4\times 10^5{\rm ~sec} \times
  (m_{3/2}/1 {\rm ~TeV})^{-3},
 \end{eqnarray}
 where $m_{3/2}$ is the gravitino mass. Notice that the gravitino mass
is $O(100{\rm ~GeV}-1{\rm ~TeV})$ in a model with gravity-mediated
supersymmetry (SUSY) breaking, resulting in a lifetime which may
affect BBN.

If the gravitino is thermally produced in the early universe, and
decays without being diluted, it completely spoils the success of
SBBN. Usually, we solve this problem by introducing inflation, which
dilutes away the primordial gravitinos. However, even with inflation,
gravitinos are produced through scattering processes of thermal
particles after reheating. The abundance
$Y_{3/2} = n_{3/2}/n_\gamma$
of the gravitino depends
on the reheating temperature $T_R$, and is given by~\cite{KM1}
 \begin{eqnarray}
  Y_{3/2} \simeq 3\times 10^{-11} \times (T_R/10^{10}{\rm GeV}). 
 \end{eqnarray}
Therefore, if the reheating temperature is too high, then gravitinos
are overproduced, and too many light nuclei are photodissociated.

We can transform our constraints on $(\tau_X, m_X Y_X)$ into constraints
on $(m_{3/2}, T_R)$. In particular, we use the projected 95\%
C.L. boundaries from Figs.~12 and 16.  For several values of the
gravitino mass, we read off the most conservative upper bound on the
reheating temperature from Fig.~17, and the results are given by
 \begin{eqnarray}
  m_{3/2}=100{\rm ~GeV} ~~~ 
  (\tau_{3/2}\simeq 4\times 10^8{\rm ~sec}) &:& 
  T_R \lesssim 2\times 10^6 {\rm ~GeV},
 \nonumber \\
  m_{3/2}=1{\rm ~TeV} ~~~ 
  (\tau_{3/2}\simeq 4\times 10^5{\rm ~sec}) &:& 
  T_R \lesssim 6\times 10^8 {\rm ~GeV},
 \nonumber \\
  m_{3/2}=3{\rm ~TeV} ~~~ 
  (\tau_{3/2}\simeq 1\times 10^4{\rm ~sec}) &:& 
  T_R \lesssim 2\times 10^{11} {\rm ~GeV}.
 \nonumber
 \end{eqnarray}
 If the gravitino is heavy enough ($m_{3/2}\gtrsim 5{\rm ~TeV}$), then
its lifetime is too short to destroy even D. In this case, our only
constraint is from the overproduction of $^4$He. If the gravitino mass
is lighter, then the lifetime is long enough to destroy D, $^3$He,
or even
$^4$He. In this case, our constraint on the reheating temperature is
more severe.

Another example of our decaying particle is the lightest superparticle
in the MSSM sector, if it is heavier than the gravitino.  In
particular, if the lightest neutralino is the lightest superparticle
in the MSSM sector, then it can be a source of high-energy photons,
since it will decay into a photon and a gravitino. In this case, we
may use BBN to constrain the MSSM.

The abundance of the lightest neutralino is determined when it
freezes out of the thermal bath. The abundance
is a function of the masses of the superparticles, and it becomes
larger as the superparticles get heavier. Thus, the upper bound on
$m_X Y_X$ can be translated into an upper bound on the mass scale of the
superparticles.

In order to investigate this scenario, we consider the simplest case
where the lightest neutralino is (almost) purely bino $\tilde{B}$.
In this case,
the lightest neutralino pair-annihilates through squark and slepton
exchange. In particular, if the right-handed sleptons are the lightest
sfermions, then the dominant annihilation is
$\tilde{B} + \tilde{B}\rightarrow l^+ + l^-$. The annihilation cross section
though this process is given by~\cite{PLB230-78}
 \begin{eqnarray}
  \langle\sigma v_{\rm rel}\rangle
  = 8\pi\alpha_1^2 \langle v^2\rangle
  \left\{
  \frac{m_{\tilde{B}}^2}{(m_{\tilde{B}}^2+m_{\tilde{l}_R}^2)^2}
  - \frac{2m_{\tilde{B}}^4}{(m_{\tilde{B}}^2+m_{\tilde{l}_R}^2)^3}
  + \frac{2m_{\tilde{B}}^6}{(m_{\tilde{B}}^2+m_{\tilde{l}_R}^2)^4}
  \right\},
 \end{eqnarray}
 where $\langle v^2\rangle$ is the average velocity squared of bino,
and we added the contributions from all three generations by assuming the
right-handed sleptons are degenerate.\footnote
 {If the bino is heavier than the top quark, then the $s$-wave contribution
annihilating into top quarks becomes important. In this paper, we do
not consider this case.}
 With this annihilation cross section, the Boltzmann equation for the
number density of binos is given by
 \begin{eqnarray}
  \dot{n}_{\tilde{B}} + 3H n_{\tilde{B}} = - 2 \langle\sigma v_{\rm
  rel}\rangle (n_{\tilde{B}}^2 - (n_{\tilde{B}}^{{\rm EQ}})^2),
 \end{eqnarray}
 where $n_{\tilde{B}}^{{\rm EQ}}$ is the equilibrium number density of
binos.  The factor 2 is present because two binos annihilate into leptons
in one collision. We solved this equation and obtained the mass
density of the bino as a function of the bino mass and the
right-handed slepton mass. (For details, see {\it e.g.}
Ref.~\cite{Kolb-Turner}). Numerically, for $m_{\tilde{B}}=100$~GeV,
$m_XY_X$ ranges from $\sim 10^{-9}$~GeV to $\sim 10^{-5}$~GeV as we
vary $m_{\tilde{l}_R}$ from 100~GeV to 1~TeV. If $m_XY_X$ is in this
range, BBN is significantly affected unless the lifetime of the bino
is shorter than $10^4$ -- $10^5$~sec (see Tables~\ref{table:ll} --
\ref{table:lh}). The lifetime of the bino is given by
 \begin{eqnarray}
  \tau_{\tilde{B}} = 
  \left[\frac{1}{48\pi}
  \frac{m_{\tilde{B}}^5\cos^2\theta_{\rm W}}
  {m_{3/2}^2M_*^2}\right]^{-1}
  \simeq 7\times 10^4{\rm ~sec}\times
  \left(\frac{m_{\tilde{B}}}{100{\rm ~GeV}}\right)^{-5}
  \left(\frac{m_{3/2}}{1{\rm ~GeV}}\right)^2.
 \end{eqnarray}
 Notice that the lifetime becomes shorter as the gravitino mass
decreases; hence, too much D and $^7$Li are destroyed if the gravitino
mass is too large. Therefore, we can convert the constraints given in
Figs.~12 and 16 into upper bounds on the gravitino mass. Since the
abundance of the bino is an increasing function of the slepton mass
$m_{\tilde{l}_R}$, the upper bound on the gravitino mass is more
severe for larger slepton masses. For example, for
$m_{\tilde{B}}=100{\rm ~GeV}$, the upper bound on the gravitino mass
is shown in Fig.~18.  At some representative values of the slepton
mass, the constraint is given by
 \begin{eqnarray}
  m_{\tilde{l}_R}=100{\rm ~GeV} &:& 
  m_{3/2}\lesssim 1{\rm ~GeV},
 \nonumber \\
  m_{\tilde{l}_R}=300{\rm ~GeV} &:& 
  m_{3/2}\lesssim 700{\rm ~MeV},
 \nonumber \\
  m_{\tilde{l}_R}=1{\rm ~TeV} &:& 
  m_{3/2}\lesssim 400{\rm ~MeV}.
 \nonumber
 \end{eqnarray}
 As expected, for a larger value of the slepton mass, the primordial
abundance of the bino gets larger, and the upper bound on the gravitino
mass becomes smaller.

Another interesting source of high-energy photons is a modulus field
$\phi$. Such fields are predicted in string-inspired supergravity
theories.  A modulus field acquires
mass from SUSY breaking, so we estimate its mass $m_\phi$
to be of the same order as the gravitino mass
(see for example~\cite{Carlos}).

In the early universe, the mass of the modulus field is negligible
compared to the expansion rate of the universe, so the modulus field
may sit far from the minimum of its potential. Since the only scale
parameter in supergravity is the Planck scale $M_*$, the initial
amplitude $\phi_0$ is naively expected to be of $O(M_*)$.  However,
this initial amplitude is too large; it leads to well-known problems
such as matter domination and distortion of the CMBR.
Here, we regard $\phi_0$ as a free parameter and
derive an upper bound on it.

Once the expansion rate becomes smaller than the mass of the modulus
field, the modulus field starts oscillating. During this period,
the energy density
of $\phi$ is proportional to $R^{-3}$ (where $R$ is the scale factor);
hence, its energy density behaves like that of non-relativistic
matter. The modulus eventually decays, when the expansion rate
becomes comparable to its decay rate. Without entropy production
from another source, the modulus density at the decay time is approximately
 \begin{eqnarray}
  m_\phi Y_{\phi} = 
  \frac{\rho_\phi}{n_\gamma} \sim 5\times 10^{10} {\rm ~GeV}
  \times (m_\phi/1{\rm ~TeV})^{1/2} (\phi_0/M_*)^2, 
 \end{eqnarray}
 where $\rho_\phi$ is the energy density of the modulus field. As in our
other models, we can convert our constraints on $(\tau_X, m_X Y_X)$
(Figs.~12, and~16) into constraints on $(m_{\phi}, \phi_0)$.
Using the most conservative of these constraints, we still obtain
very stringent bounds on the initial amplitude of the modulus
field $\phi_0$:
 \begin{eqnarray}
  m_{\phi}=100{\rm ~GeV} ~~~ (\tau_{\phi}\sim 4 \times 10^8{\rm ~sec}) &:&
  \phi_0 \lesssim 1 \times 10^7 {\rm ~GeV},
 \nonumber \\
  m_{\phi}=1{\rm ~TeV} ~~~ (\tau_{\phi}\sim 4 \times 10^5{\rm ~sec}) &:&
  \phi_0 \lesssim 5 \times 10^8 {\rm ~GeV},
 \nonumber \\
  m_{\phi}=3 {\rm ~TeV} ~~~ (\tau_{\phi}\sim 1\times 10^4{\rm ~sec}) &:&
  \phi_0 \lesssim 9 \times 10^{9} {\rm ~GeV}.
 \nonumber
 \end{eqnarray}
 Clearly, our upper bound from BBN rules out our naive expectation
that $\phi_0 \sim M_*$. It is important to notice that (conventional)
inflation cannot solve this difficulty by diluting the coherent mode
of the modulus field. This is because the expansion rate of the
universe is usually much larger than the mass of the modulus field,
and hence the modulus field does not start oscillation. One attractive
solution is a thermal inflation model proposed by Lyth and
Stewart~\cite{Lyth-Stewart}. In the thermal inflation model, a
mini-inflation of about $\sim$ 10 $e$-folds reduces the modulus
density. Even if thermal inflation occurs, there may remain a
significant modulus energy density, which decays to high-energy
photons. Thus, BBN gives a stringent constraint on the thermal
inflation model.

\section{Discussion and Conclusions}
\label{sec:summary}

We have discussed the photodissociation of light elements due to the
radiative decay of a massive particle, and we have shown how we can
constrain our model parameters from the observed light-element
abundances.  We adopted both
low and high $^4$He values in this paper, and
we obtained constraints on the properties of the radiatively decaying
particle in each case.

When we adopt the low $^4$He value, we find that a non-vanishing
amount of such a long-lived, massive particle is preferred: $m_X Y_X
\gtrsim 10 ^ {-10} {\rm GeV}$ and $10^4 {\rm sec} \lesssim \tau_X
\lesssim 10^6 {\rm sec}$. On the other hand, consistency with the
observations imposes upper bounds on $m_X Y_X$ in each cases.

We have also studied the photodissociation of $^7$Li and $^6$Li in
this paper. These processes do not affect the D, $^3$He,
and $^4$He abundances,
because $^7$Li and $^6$Li are many orders of magnitude less abundant
than D,$^3$He, and $^4$He.
When we examine the region of parameter space where
the predicted abundances agree well with the observed $^7$Li and the
low $^4$He observations, we find that the produced $^6$Li/H may be of
order $10^{-12}$, which is two orders of magnitude larger than the
prediction of SBBN (see Figs.~7 and~13). The predicted $^6$Li is
consistent with the observed upper bound Eq.~(\ref{Li6}) throughout
the region of parameter space we are interested in. Although presently
it is believed that the observed $^6$Li abundance is produced by
spallation, our model suggests another origin: the observed $^6$Li may
be produced by the photodissociation of $^7$Li.

We have also discussed candidates for our radiatively decaying
particle. Our first example is the gravitino. In this case, we can
constrain the reheating temperature after inflation, because it determines
the abundance of the gravitino. We obtained the stringent bounds
$T_R \lesssim
10^8 {\rm GeV} - 10^9 {\rm GeV}
$ 
for $100{\rm ~GeV}
\lesssim  m_{3/2} 
\lesssim 1{\rm ~TeV}$.
Our second example is the lightest neutralino which is heavier than
the gravitino. When the neutralino is the lightest superparticle in
the MSSM sector, it can decay into a photon and a gravitino. If we
assume the lightest neutralino is pure bino, and its mass is about 100
GeV, then the relic number density of binos is related to the right-handed
slepton mass, because they annihilate mainly through right-handed
slepton exchange. For this case, we obtained an upper bound on the
gravitino mass:  $m_{3/2} \lesssim 400 {\rm ~MeV} - 1 {\rm ~GeV}$ for
$100 {\rm ~GeV} \lesssim m_{\tilde{l}_R} \lesssim 1 {\rm ~TeV}$.  Our
third example is a modulus field. We obtained a severe constraint on
its initial amplitude, $ \phi_0 \lesssim 10^8 {\rm ~GeV} - 10^{9}
{\rm ~GeV}$ for $100{\rm ~GeV} \lesssim m_{3/2} \lesssim 1{\rm
~TeV}$. This bound is well below the Planck scale, so it suggests the
need for a dilution mechanism, such as thermal inflation.

\section*{Acknowledgement}

This work was supported by the Director, Office of Energy
Research, Office of Basic Energy Services, of the U.S.
Department of Energy under Contract DE-AC03-76SF00098. K.K. is
supported by JSPS Research Fellowship for Young Scientists.

\appendix

\section*{}
\label{app:analysis}

In this appendix, we explain how we answer the question,
``How well does our simulation of BBN agree with the observed
light-element abundances?'' To be more precise, we rephrase
the question as, ``At what confidence level is our simulation
of BBN excluded by the observed light-element abundances?''

From our Monte-Carlo BBN simulation, we obtain the theoretical
probability density function (p.d.f.)  $p_{247}^{th}(y_2^{th}, Y^{th},
\log_{10}y_7^{th})$ of our simulated light-element abundances
$y_2^{th}, Y^{th},$ and $\log_{10}y_7^{th}$.  We find that
$p_{247}^{th}(y_2^{th}, Y^{th}, \log_{10}y_7^{th})$ is
well-approximated by a multivariate Gaussian distribution function:
\beq
p_{247}^{th} (y^{th}_2,Y^{th}, \log_{10}y^{th}_7)
 = p^{Gauss}_2(         y^{th}_2) \times
   p^{Gauss}_4(         Y^{th}  ) \times
p^{Gauss}_7(\log_{10}y^{th}_7),
\eeq
where
\beq
p^{Gauss}(x;\bar{x},\sigma) &=&
 \frac{1}{\sqrt{2 \pi}\sigma}
 \exp \left[ -\frac{1}{2}
            \left( \frac{ x-\bar{x} }{\sigma} \right)^2
      \right].
\eeq

Note that
$p_{247}^{th}(y_2^{th}, Y^{th}, \log_{10}y_7^{th})$ depends upon the
parameters {\bf p} $ = (\eta,...)$
of our theory.
(The ellipses refer to parameters in non-standard BBN, {\it
e.g.}, $m_X Y_X$, $\tau_X$.)  In particular, the means and standard
deviations of $p_{247}^{th}(y_2^{th}, Y^{th}, \log_{10}y_7^{th})$ are
functions of {\bf p}.

For BBN with a radiatively decaying particle, we also consider
the ratio $r^{th} = y_3^{th} / y_2^{th}$.
Our Monte-Carlo BBN simulation allows us to find the
p.d.f. $p_{23}^{th}(r^{th})$.  We
approximate $p_{23}^{th}$ by a Gaussian, and neglect the correlation
between $r$ and $y_2^{th}, Y^{th}, \log_{10}y_7^{th}$.  This
assumption (which is justified in work to be published by one of
the authors~\cite{holtmann}), allows us to write
\beq
p_{2347}^{th}(y_2^{th}, Y^{th}, r^{th}, \log_{10} y_7^{th}; {\bf p})
 = p_{23}^{th}(r^{th}; {\bf p}) \times
   p_{247}^{th}(y_2^{th}, Y^{th}, \log_{10} y_7^{th}; {\bf p}).
\eeq

We want to compare these theoretical calculations to the
observed light-element abundances
$y^{obs}_2, Y^{obs},$ and $\log_{10}y^{obs}_7$.
Since the observations of the light-element abundances
are independent, we can factor the p.d.f.
$p_{247}^{obs}(y_2^{obs}, Y^{obs}, \log_{10} y_7^{obs})$ as
 \beq
  p_{247}^{obs}(y_2^{obs}, Y^{obs}, \log_{10} y_7^{obs})
  &=& p^{obs}_2(y^{obs}_2) \times
      p^{obs}_4(Y^{obs}) \times
      p^{obs}_7(\log_{10}y^{obs}_7).
 \eeq
 We assume Gaussian p.d.f.'s for $y^{obs}_2, Y^{obs}$, and
$\log_{10}y^{obs}_7$.  We use the mean abundances and standard
deviations given in Equations~(\ref{lowD}),
(\ref{lowHe}), (\ref{highHe}), and (\ref{Li7}).  Since we
have two discordant values of $^4$He,
we considered two cases, {\it i.e.}, high and low values of $^4$He
abundances. 

$^3$He is more complicated.  Aside from the trivial positivity requirement,
we only have an upper bound on $r^{obs}$
(the primordial $^3$He/D, as deduced from solar-system observations);
namely, $r^{obs} < r_{\odot}$.
Because of this, we choose for the p.d.f.
\beq
  p_{23}^{obs}(r^{obs}) = \left\{ \begin{array}{ll}
                     0  & , \ r^{obs} < 0 \\
                     N  & , \ 0 < r^{obs} < \overline{r_\odot} \\
                     N  \exp \left( -\frac{1}{2}
                             \left( \frac{r^{obs}-\overline{r_\odot}}
                                        {\sigma_{r \odot}}
                             \right)  ^2
                        \right)
                        & , \ \overline{r_\odot} < r^{obs}
                     \end{array}
                     \right.
  \label{eqn:p23}
\eeq
where the normalization factor is
$N = 1/(\overline{r_\odot}  + \sigma_{r \odot} \sqrt{\pi/2})$,
and $ \overline{r_\odot}, \sigma_{r \odot} $ are given in Eq. (\ref{He3/D}).

Since the observations of the light-element abundances
are independent, we can write the total observational p.d.f. for BBN$+X$ as
\beq
p^{obs}_{2347}(y_2^{obs}, Y^{obs}, r^{obs}, \log_{10} y_7^{obs})
 = p_{23}^{obs}(r^{obs})
  \times p^{obs}_{247}(y_2^{obs}, Y^{obs}, \log_{10} y_7^{obs}).
\eeq
To simplify the notation, we write
\beq
  {\bf a}^{th}  &=& (y_2^{th} , Y^{th} , r^{th} , \log_{10} y_7^{th})  \\
  {\bf a}^{obs} &=& (y_2^{obs}, Y^{obs}, r^{obs}, \log_{10} y_7^{obs}) \\
  {\bf b}^{th}  &=& (y_2^{th} , Y^{th} , \log_{10} y_7^{th})  \\
  {\bf b}^{obs} &=& (y_2^{obs}, Y^{obs}, \log_{10} y_7^{obs}).
\eeq
Then the quantities $\Delta{\bf a} = {\bf a}^{th} - {\bf a}^{obs}$
have a p.d.f. given by:
\beq
p_{247}^\Delta(\Delta{\bf b})
  &=&   \int d{\bf b}^{obs}~p_{247}^{obs}({\bf b}^{obs})
        ~\int~d{\bf b}^{th}~p_{247}^{th}({\bf b}^{th})
         \delta(\Delta{\bf b} - ({\bf b}^{th} - {\bf b}^{obs}))
      \nonumber \\
  &=& \int d{\bf b}~p_{247}^{th}({\bf b})
                    p_{247}^{obs}({\bf b} - \Delta{\bf b}),
   ~~~\mbox{[SBBN]}    \label{pdf-SBBN}  \\
p_{2347}^\Delta(\Delta{\bf a})
  &=&   \int d{\bf a}^{obs}~p_{2347}^{obs}({\bf a}^{obs})
        ~\int~d{\bf a}^{th}~p_{2347}^{th}({\bf a}^{th})
         \delta(\Delta{\bf a} - ({\bf a}^{th} - {\bf a}^{obs}))
      \nonumber \\
  &=& \int d{\bf a}~p_{2347}^{th}({\bf a})
                    p_{2347}^{obs}({\bf a} - \Delta{\bf a}),
   ~~~\mbox{[BBN$+X$]}  \label{pdf-BBNX}
\eeq
where we have suppressed the dependence of
$p_{247}^\Delta(\Delta{\bf b})$, $p_{247}^{th}({\bf b}^{th})$,
$p_{2347}^\Delta(\Delta{\bf a})$, and $p_{2347}^{th}({\bf a}^{th})$
on the theory parameters {\bf p}.
The integral in Eq. (\ref{pdf-SBBN})
is simply a Gaussian:
\beq
    p_{247}^\Delta(\Delta{\bf b})
   &=& \prod_{i=2,4,7} \frac{1}{\sqrt{2 \pi} \sigma_i}
        \exp\left[-\frac12 \left(
                   \frac{\Delta b_i - \overline{\Delta b_i}}
                        {\sigma_i^2} \right) ^2\right],
\eeq
where $\overline{\Delta b_i} = \overline{b^{th}_i}-\overline{b^{obs}_i}$, $\sigma^2_i =
(\sigma^{th}_i)^2 + (\sigma^{obs}_i)^2$, and $i$ runs over $y_2$, $Y$ and 
log$_{10}y_7$.
To evaluate Eq. (\ref{pdf-BBNX}), we note that
\beq
  p_{2347}^{obs}({\bf a}^{obs})
   = p_{23}(r^{obs}) \times p_{247}^{obs}({\bf b}^{obs})     \\
  p_{2347}^{th} ({\bf a}^{th} )
   = p_{23}(r^{th} ) \times p_{247}^{th} ({\bf b}^{th} ),
\eeq
where $p_{247}^{obs}$ and $p_{247}^{th}$ are Gaussian.
The integral then factors as
\beq
    \label{gpdf}
    p_{2347}^\Delta(\Delta{\bf a})
    &=& \int dr p_{23}^{th}(r) p_{23}^{obs}(r - \Delta r)
        \times
        \int d{\bf b}~p_{247}^{th}({\bf b})
                    p_{247}^{obs}({\bf b} - \Delta{\bf b})   \\
    &=& p_{23}^\Delta(\Delta r)
        \times
        p_{247}^\Delta(\Delta{\bf b}).
\eeq
The first integral can be evaluated as
\beq
     p_{23}^\Delta(\Delta r)
   &=& \int_{-\infty}^{\infty} dr p_{23}^{th}(r + \Delta r) p_{23}^{obs}(r)     \\
   &=& \int_{0}^{\overline{r_\odot}} dr \frac{N}{\sqrt{2 \pi} \sigma_r^{th}}
      \exp \left[ -\frac{1}{2} \left( \frac{ r + \Delta r - \overline{r^{th}}}
          {\sigma_r^{th}} \right)^2 \right] \nonumber \\
   & & + \int_{\overline{r_\odot}}^{\infty} dr 
   \frac{N}{\sqrt{2 \pi} \sigma_r^{th}}
   \exp \left( -\frac{1}{2} \left[
   \left( \frac{r + \Delta r - \overline{r^{th}}}{\sigma_r^{th}} \right)^2
 + \left( \frac{r - \overline{r_\odot}} {\sigma_{r \odot}} \right)^2
    \right] \right) \\
   &=& \frac{N}{2} \left[
        {\rm erf} \left( \frac{\overline{r_\odot} + \Delta r
          - \overline{r^{th}}}{\sqrt{2}\sigma_r^{th}} \right)
      - {\rm erf} \left( \frac{ \Delta r
          - \overline{r^{th}}}{\sqrt{2}\sigma_r^{th}} \right)  \right]
       \nonumber \\
   & & + \frac{N}{2} \frac{\sigma_{r \odot}}{\sigma_r}
      \left[ 1 -  {\rm erf} \left( \frac{\overline{r_\odot} - \bar{r}}
          {\sqrt{2}\tilde{\sigma_r}} \right) \right]
      \exp \left[-\frac{1}{2} \left(\frac{ \Delta r - \overline{r^{th}}
           + \overline{r_\odot} }
          {\sigma_r} \right)^2 \right],
\eeq
where
$\tilde{\sigma_r}^{-2} = (\sigma_r^{th})^{-2} + (\sigma_{r \odot})^{-2}$,
$\bar{r} / \tilde{\sigma_r}^2 = \overline{r_\odot}/(\sigma_{r \odot})^2
-(\Delta r - \overline{r^{th}})/(\sigma_r^{th})^2$,
and $\sigma_r^2 = (\sigma_r^{th})^2 + (\sigma_{r \odot})^2$.

Our question can now be rephrased as, ``At what confidence level is
$\Delta{\bf a} = 0$ excluded?''  To answer this,
we need to consider the region $S$ in abundance space where the
value of the p.d.f. is higher than
\beq
\tilde{p} = \left\{ \begin{array}{cc}
             p_{247}^\Delta(\Delta{\bf b} = 0;{\bf p}) ~~~ \mbox{[SBBN]}    \\
             p_{2347}^\Delta(\Delta{\bf a} = 0;{\bf p}) ~~~\mbox{[BBN$+X$]}  \\
             \end{array} \right.
\eeq
Mathematically phrased,
\beq
{\rm C.L.}({\bf p})
 = \int_S d(\Delta{\bf b})
          p_{247}^\Delta(\Delta{\bf b}; {\bf p}),  ~~~\mbox{[SBBN]}    \\
{\rm C.L.}({\bf p})
 = \int_S d(\Delta{\bf a})
          p_{2347}^\Delta(\Delta{\bf a}; {\bf p}), ~~~\mbox{[BBN$+X$]}
  \label{CL-BBNX}
\eeq
where
\beq
S &=& \{ \Delta{\bf b} :
    p_{247}^\Delta(\Delta{\bf b};{\bf p}) > \tilde{p} \},
   ~~~\mbox{[SBBN]},  \\
S &=& \{ \Delta{\bf a} :
    p_{2347}^\Delta(\Delta{\bf a};{\bf p}) > \tilde{p} \},
   ~~~\mbox{[BBN$+X$]}.
\eeq
We use the C.L. to constrain various scenarios of BBN.

In the SBBN case, the integral is Gaussian and is easily expressed in terms
of
$\chi^2 = -2 \log[(2\pi)^{3/2}
          \sigma_2 \sigma_4 \sigma_7
          \tilde{p}]$
(See Eqs.~(\ref{chi2}) and~(\ref{CL})):
\beq
{\rm C.L.}({\bf p})
  =
  -\sqrt{\frac{2 \chi^2}{\pi}}
  \exp \left( -\frac{\chi^2}{2} \right)
  + {\rm erf} \left( \sqrt{\frac{\chi^2}{2}} \right),
  ~~~\mbox{[SBBN]}.
\eeq
To evaluate the BBN$+X$ integral Eq. (\ref{CL-BBNX}),
we separate the Gaussian variables
$\Delta {\bf b}$
from the non-Gaussian variable $\Delta r$,
using the decomposition in Eq.~(\ref{gpdf}):
\beq
  {\rm C.L.}({\bf p})
  = \int_{S_{23}} d(\Delta r) p_{23}^\Delta(\Delta r)
    \int_{S_{247}(\Delta r)} d(\Delta {\bf b})
                             p_{247}^\Delta(\Delta {\bf b}),
\eeq
where $S_{23}$ = the set of $\Delta r$ such that
$p_{23}^\Delta(\Delta r) \geq \tilde{p} / {\rm max}(p_{247})$;
and
$S_{247}(\Delta r)$ = the set of $\Delta {\bf b}$ such that
$p_{247}^\Delta(\Delta {\bf b}) \geq \tilde{p} / p_{23}^\Delta(\Delta r)$.
We can easily evaluate the Gaussian integral.  Again,
the result is conveniently expressed in terms of $\chi^2$.
\beq
{\rm C.L.}({\bf p})
  = \int_{S_{23}} d(\Delta r) p_{23}^\Delta(\Delta r)
   \left[ -\sqrt{\frac{2}{\pi}} \chi(\Delta r) e^{-\chi^2(\Delta r)/2}
   + \rm{erf} \left( \frac{\chi(\Delta r)}{\sqrt{2}} \right) \right],
  \label{CL-int-dr}
\eeq
where
$\chi(\Delta r) = \sqrt{-2 \log[(2\pi)^{3/2}
                  \sigma_2 \sigma_4 \sigma_7
                  \tilde{p}/p_{23}(\Delta r)]}$.
We then evaluate Eq. (\ref{CL-int-dr}) numerically.

Our confidence level is calculated for four degrees of freedom
$\Delta a_i$ (three, in the case of SBBN).
It denotes our certainty that a given point {\bf p} in
the parameter space of the theory is excluded by the observed
abundances. In order to compare our theory with a late-decaying
particle (three parameters {\bf p}:
$\tau_X, m_X Y_X$, and $\eta$) to a
theory with a different number of parameters ({\it e.g.},
only one in SBBN), one
would want to use a $\chi^2$ variable in these parameters.
This transformation would be possible
if the abundances $a_i$ were linear in the theory
parameters {\bf p}.
In this case, we could integrate out a theory parameter
such as $\eta$ and set a C.L. exclusion limit (with a reduced number
of degrees of freedom) on the remaining parameters. However, the
$a_i$ turn out to be highly non-linear functions of {\bf p},
so integrating out a theory parameter turns out
to have little meaning.  Instead, we project out various theory
parameters (as explained in Section~\ref{sec:compar}) to present our results
as graphs.


\newpage

\begin{table}[t]
\begin{center}
\begin{tabular}{lcc}
  & $N_\nu$ (95 $\%$ C.L.) & $\eta \times 10^{10}$ (95 $\%$ C.L.)\\
\hline
Low $^4$He &
   2.1$^{+1.0}_{-0.8}$ &
   4.7$^{+1.0}_{-0.8}$ \\
High $^4$He &
     2.8$^{+1.0}_{-1.0}$ &
     5.0$^{+1.0}_{-0.8}$ \\
\end{tabular}
\caption{Observational constraints on $\eta$ and $N_\nu$ in SBBN}
\label{table:nueta}
\end{center}
\end{table}

\begin{table}[t]
\begin{center}
\begin{tabular}{rlrrr}
& {Photofission Reactions} & 1$\sigma$ Uncertainty & Threshold Energy & Ref.\\
\hline
 1. &   ${\rm D} + \gamma \rightarrow p + n$
                      &  6\% &  2.2 MeV & \cite{Evans}\\
 2. &   ${\rm T} + \gamma \rightarrow n + {\rm D}$
                      & 14\% &  6.3 MeV & \cite{ZP208-129,PRL44-129}\\
 3. &   ${\rm T} + \gamma \rightarrow p + 2n$
                      &  7\% &  8.5 MeV & \cite{PRL44-129} \\
 4. &$^3{\rm He} + \gamma \rightarrow p + {\rm D}$
                      & 10\% &  5.5 MeV & \cite{PL11-137} \\
 5. &$^3{\rm He} + \gamma \rightarrow n + 2p
                    $ & 15\% &  7.7 MeV & \cite{PL11-137} \\
 6. &$^4{\rm He} + \gamma \rightarrow p + {\rm T}$
                      &  4\% & 19.8 MeV & \cite{PL11-137} \\
 7. &$^4{\rm He} + \gamma \rightarrow n +~^3{\rm He}$
                      &  5\% & 20.6 MeV & \cite{CJP53-802,PLB47-433} \\
 8. &$^4{\rm He} + \gamma \rightarrow p + n + {\rm D}$
                      & 14\% & 26.1 MeV & \cite{SJNP19-598} \\
 9. &$^6{\rm Li} + \gamma \rightarrow {\rm anything}$
                      &  4\% &  5.7 MeV & \cite{Berman} \\
10. &$^7{\rm Li} + \gamma \rightarrow 2n + {\rm anything}$
                      &  9\% & 10.9 MeV & \cite{Berman} \\
11. &$^7{\rm Li} + \gamma \rightarrow n +~^6{\rm Li}$
                      &  4\% &  7.2 MeV & \cite{Berman} \\
12. &$^7{\rm Li} + \gamma \rightarrow~^4{\rm He} + {\rm anything}$
                      &  9\% &  2.5 MeV & \cite{Berman} \\
13. &$^7{\rm Be} + \gamma \rightarrow p +~^6{\rm Li}$
                      &      &          & \\
14. &$^7{\rm Be} + \gamma \rightarrow~{\rm anything~except}~^6{\rm Li}$
                      &      &          & \\
\end{tabular}
\caption{Photodissociation processes, and the 1-$\sigma$ uncertainty
         in the cross sections.
         Since there is no experimental data on photodissociation
         of $^7$Be, we assume in this paper that the rate for Reaction 13
         is the same as for Reaction 11, and the rate for Reaction 14 is
         the sum of the rates for Reactions 10 and 12.}
\label{table:pf}
\end{center}
\end{table}

\newpage

\begin{table}[t]
\begin{center}
\begin{tabular}{ccccccc}
 & $\tau_X=10^4$ sec & 10$^5$ sec & 10$^6$ sec & 10$^7$  sec &
 10$^8$ sec & 10$^9$ sec \\ \hline
 95\% C.L.&
 6$\times$10$^{-6}$&
 9$\times$10$^{-9}$&
 1$\times$10$^{-9}$&
 3$\times$10$^{-12}$&
 4$\times$10$^{-13}$&
 3$\times$10$^{-13}$\\
 68\% C.L.&
 (6 -- 60)$\times$10$^{-7}$&
 (8 -- 80)$\times$10$^{-10}$&
 (1 -- 9)$\times$10$^{-10}$&
 & & \\
 \end{tabular}
 \caption{
Upper or (upper -- lower) bound on $m_XY_X$ in units of GeV
  for the case of low $^4$He.  Note that the C.L. is for four
  degrees of freedom, and $\eta$ is varied to give the extreme values
  for $m_XY_X$.
}
 \label{table:ll}
 \end{center}
 \end{table}

\begin{table}[t]
\begin{center}
\begin{tabular}{ccccccc} 
 & $\tau_X=10^4$ sec & 10$^5$ sec & 10$^6$ sec & 10$^7$  sec &
10$^8$ sec & 10$^9$ sec \\ \hline
95\% C.L.&
7$\times$ 10$^{-6}$&
7$\times$ 10$^{-9}$&
8$\times$ 10$^{-10}$&
6$\times$ 10$^{-12}$&
7$\times$ 10$^{-13}$&
5$\times$ 10$^{-13}$ \\ 
68\% C.L.&
5$\times$ 10$^{-6}$&
5$\times$ 10$^{-9}$&
6$\times$ 10$^{-10}$&
3$\times$ 10$^{-12}$&
5$\times$ 10$^{-13}$&
3$\times$ 10$^{-13}$ \\ 
\end{tabular}
\caption{Same as Table~\protect\ref{table:ll}, 
except for high $^4$He.}
\label{table:lh}
\end{center}
\end{table}

%
\begin{figure}[hp]
\begin{center}
\vspace{2cm}
    \epsfxsize=1.0\textwidth\epsfbox{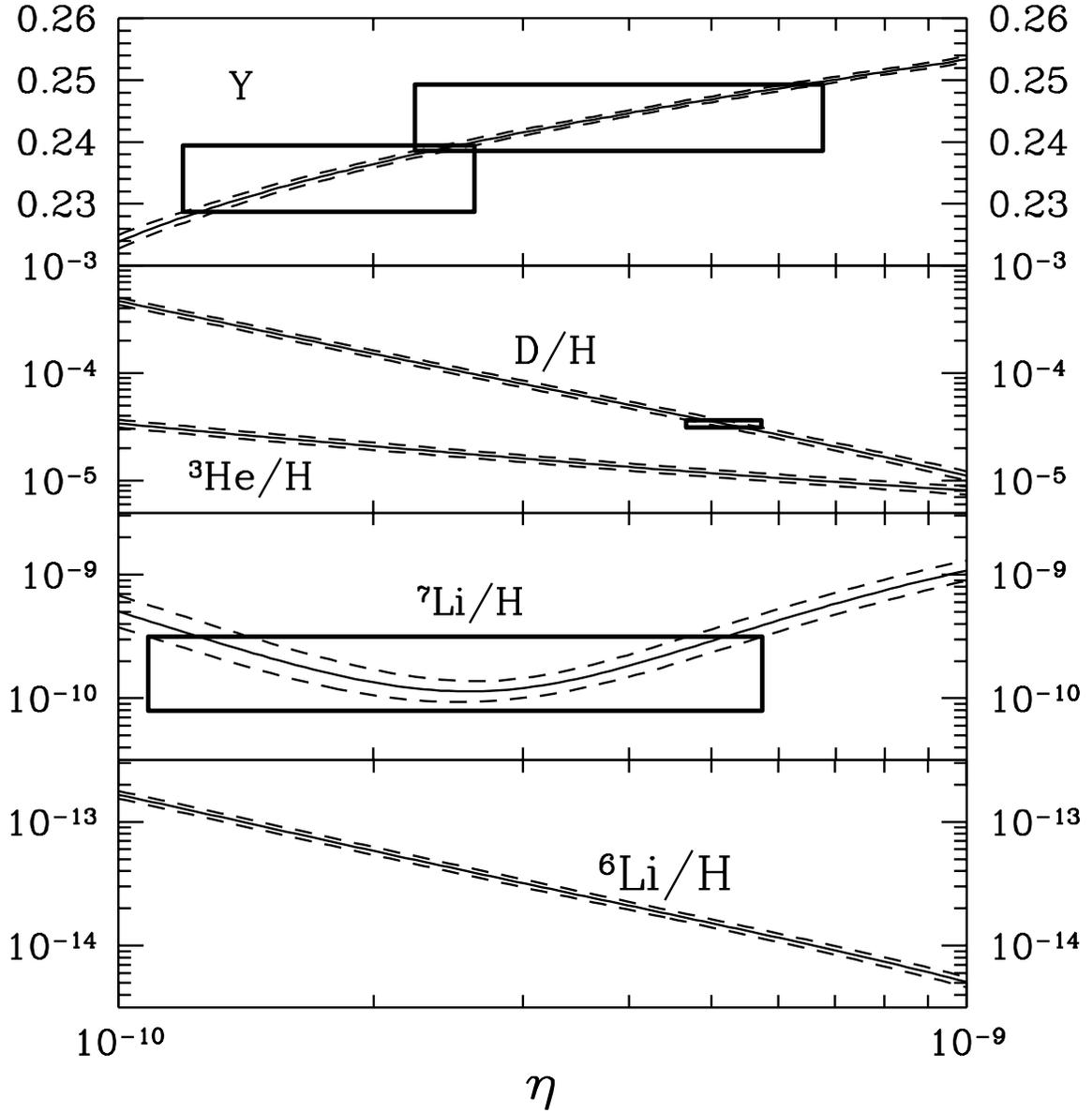}
\vspace{1cm}
\caption{
SBBN prediction of the abundances of the light
elements. The solid lines are the central values of the predictions, and the
dotted lines represents the 1-$\sigma$ uncertainties. The boxes denote 
the 1-$\sigma$ observational constraints.
}
\label{fig:sbbn}
\end{center}
\end{figure}
%

\newpage

%
\begin{figure}[hp]
\begin{center}
\vspace{3cm}
    \epsfxsize=1.0\textwidth\epsfbox{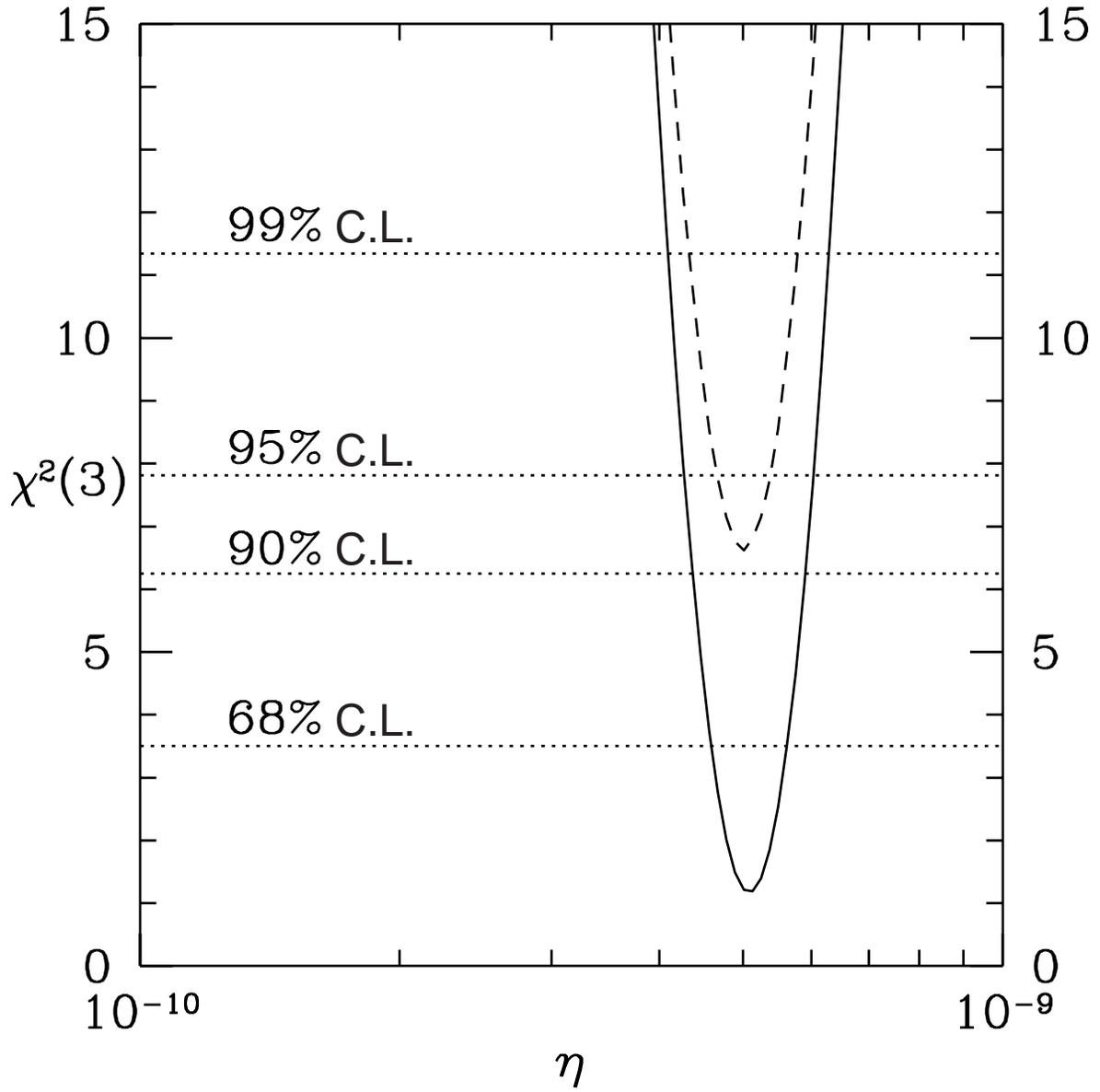}
\vspace{1cm}
\caption{
$\chi^2$ as a function of $\eta$, for SBBN with three degrees of
freedom $(\eta, \tau_X, m_X Y_X)$. We show our results for both of the
$^4$He abundances deduced from observation: low  $^4$He (dashed), high
$^4$He (solid).
}
 \label{fig:sbbnchi2}
\end{center}
\end{figure}
%

\newpage

%
\begin{figure}[hp]
\begin{center}
\vspace{6cm}
    \epsfxsize=0.6\textwidth\epsfbox{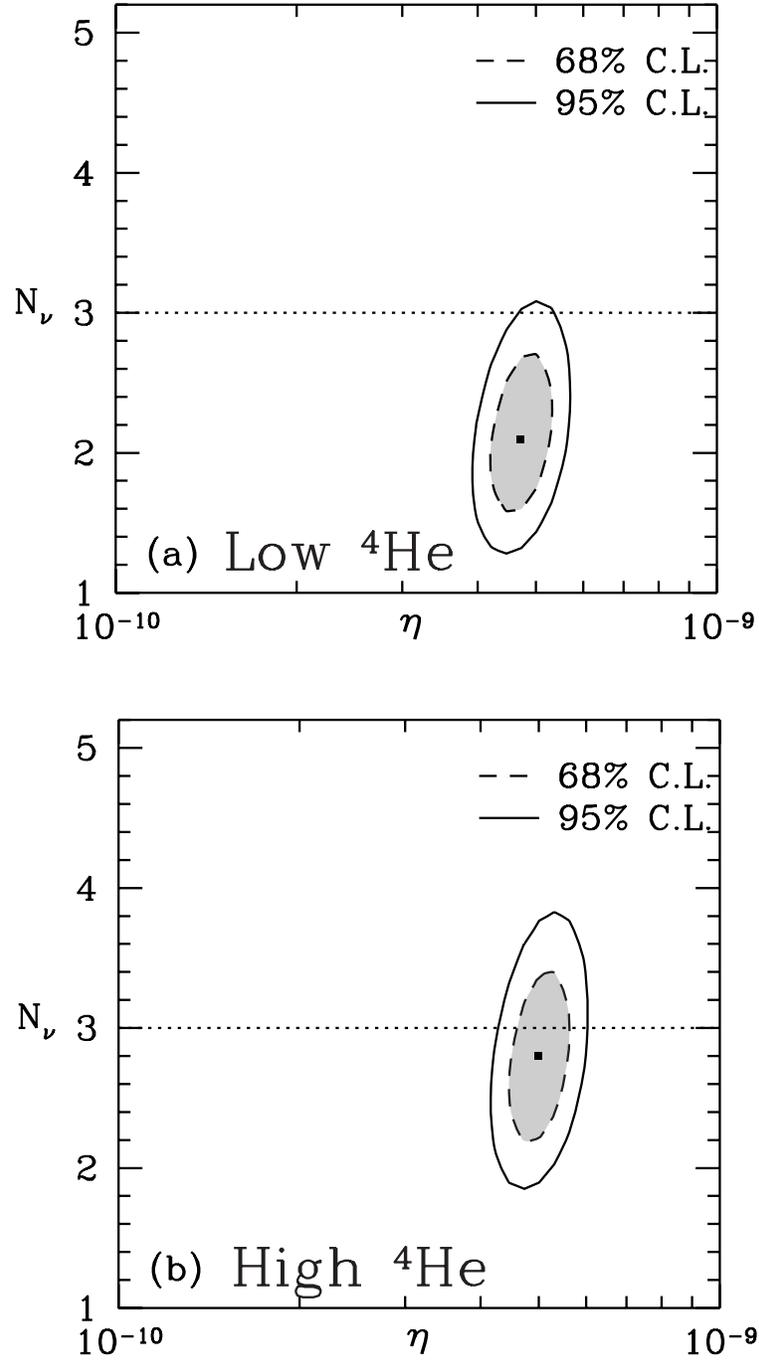}
\vspace{1cm}
\caption{
Figure 3: C.L. for BBN as a function of $\eta$ and $N_\nu$, with
(a) low value of $Y$, and (b) high value of $Y$. The filled square
denotes the best-fit point.
}
 \label{fig:nnu}
\end{center}
\end{figure}
%

\newpage

%
\begin{figure}[hp]
\begin{center}
\vspace{2cm}
    \epsfxsize=1.0\textwidth\epsfbox{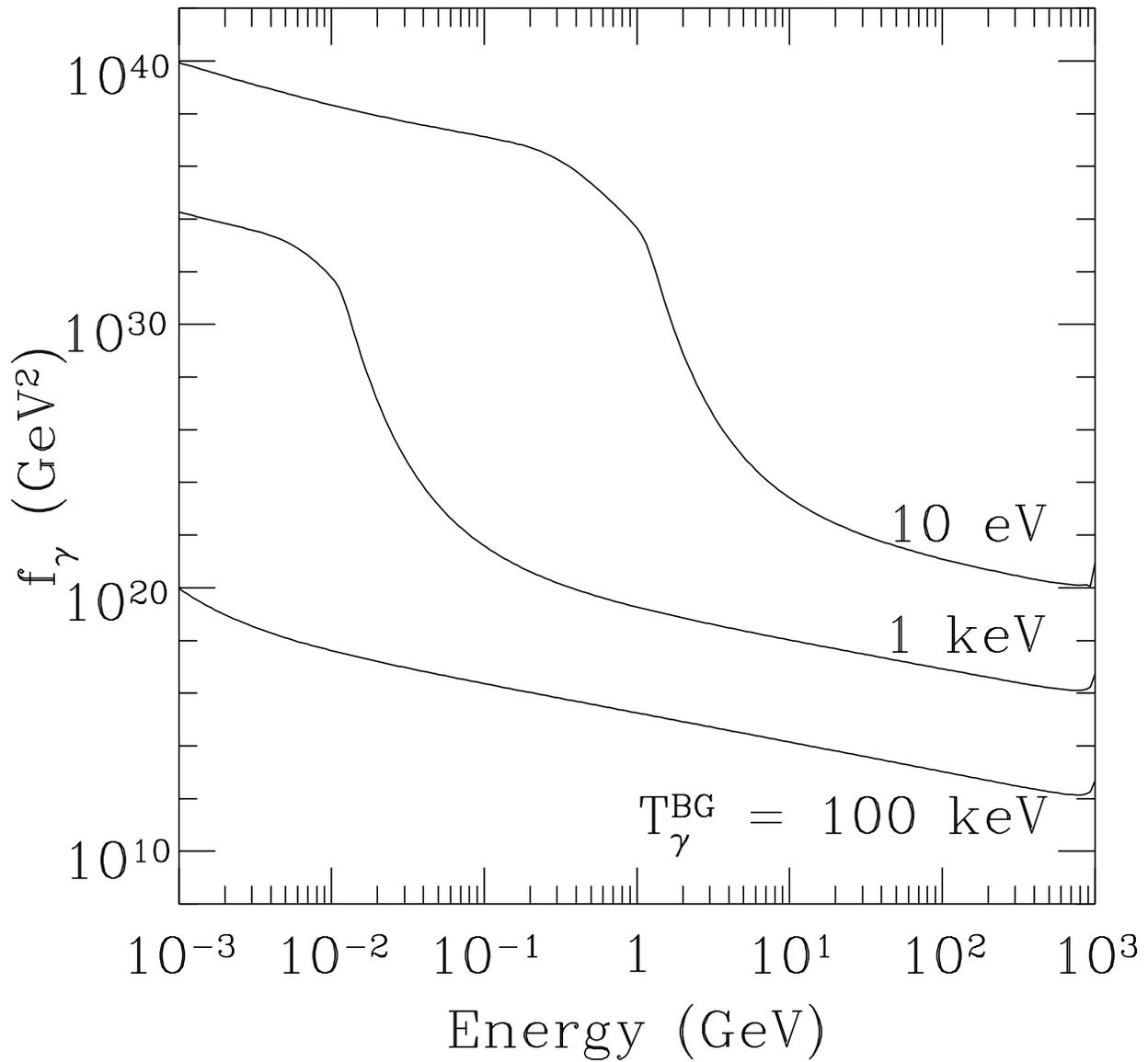}
\vspace{1cm}
\caption{
Photon spectrum $f_\gamma = dn_\gamma/dE_\gamma$ for several
background temperatures $T_{\gamma}^{\rm BG}$.
}
\label{fig:photon}
\end{center}
\end{figure}
%

\newpage

%
\begin{figure}[hp]
\begin{center}
\vspace{2cm}
    \epsfxsize=1.0\textwidth\epsfbox{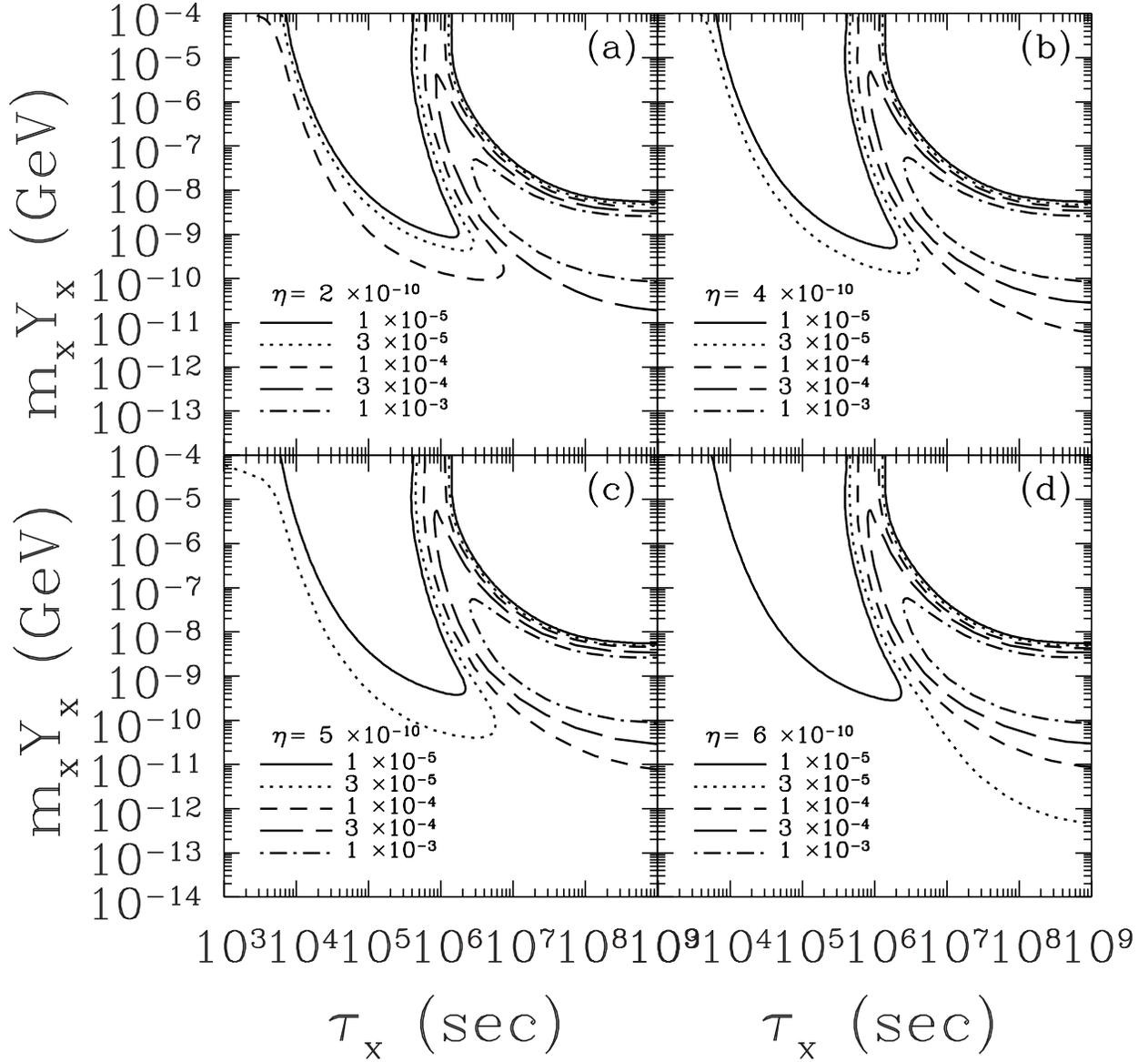}
\vspace{1cm}
\caption{
The abundance of D/H in the $m_X Y_X$ vs. $\tau_X$ plane with (a)
$\eta=2\times 10^{-10}$, (b) $\eta=4\times 10^{-10}$, (c)
$\eta=5\times 10^{-10}$, and (d) $\eta=6\times 10^{-10}$.
}
\label{fig:bbnx_d}
\end{center}
\end{figure}
%

\newpage

%
\begin{figure}[hp]
\begin{center}
\vspace{2cm}
    \epsfxsize=1.0\textwidth\epsfbox{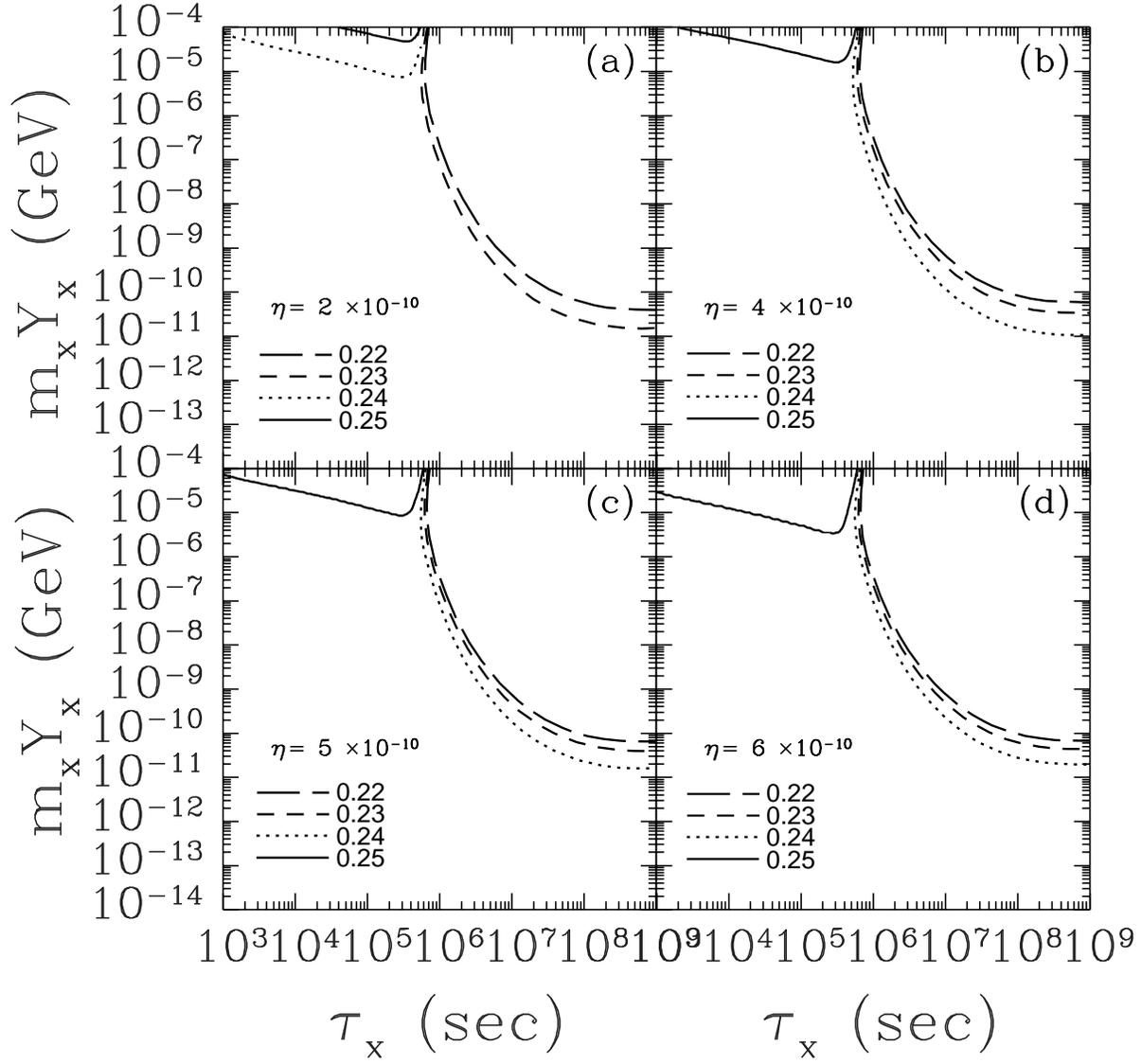}
\vspace{1cm}
\caption{
The mass fraction of $^4$He, for the same theory parameters as in
Fig.~5.
}
\label{fig:bbnx_4he}
\end{center}
\end{figure}
%

\newpage

%
\begin{figure}[hp]
\begin{center}
\vspace{2cm}
    \epsfxsize=1.0\textwidth\epsfbox{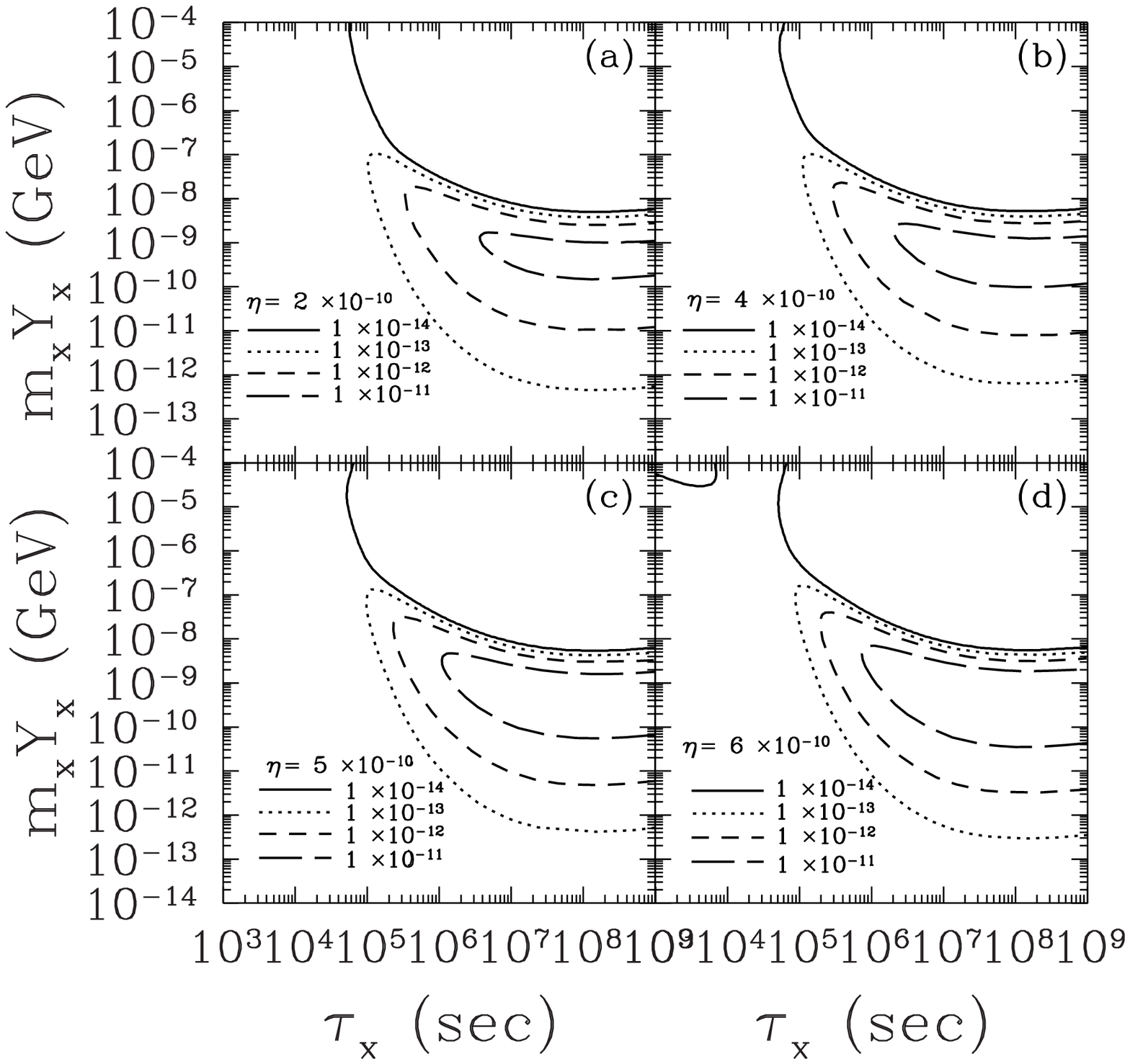}
\vspace{1cm}
\caption{
The abundance of $^6$Li/H, for the same
          theory parameters as in Fig.~5.
}
\label{fig:bbnx_6li}
\end{center}
\end{figure}
%

\newpage

%
\begin{figure}[hp]
\begin{center}
\vspace{2cm}
     \epsfxsize=1.0\textwidth\epsfbox{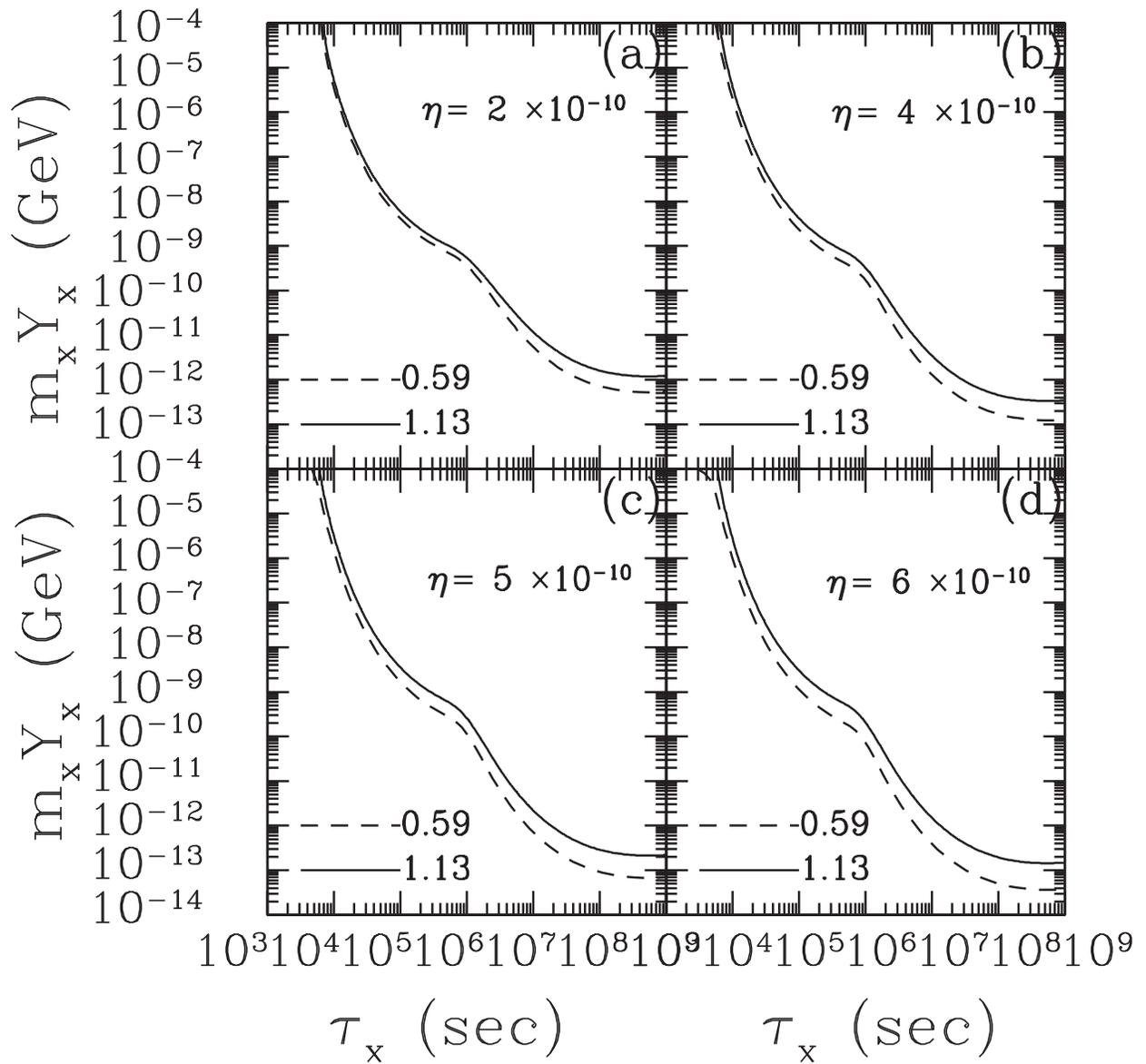}
\vspace{1cm}
\caption{
The abundance of $^3$He/D, for the same theory parameters as in Fig.~5.
}
\label{fig:bbnx_r}
\end{center}
\end{figure}
%

\newpage

%
\begin{figure}[hp]
\begin{center}
\vspace{2cm}
     \epsfxsize=1.0\textwidth\epsfbox{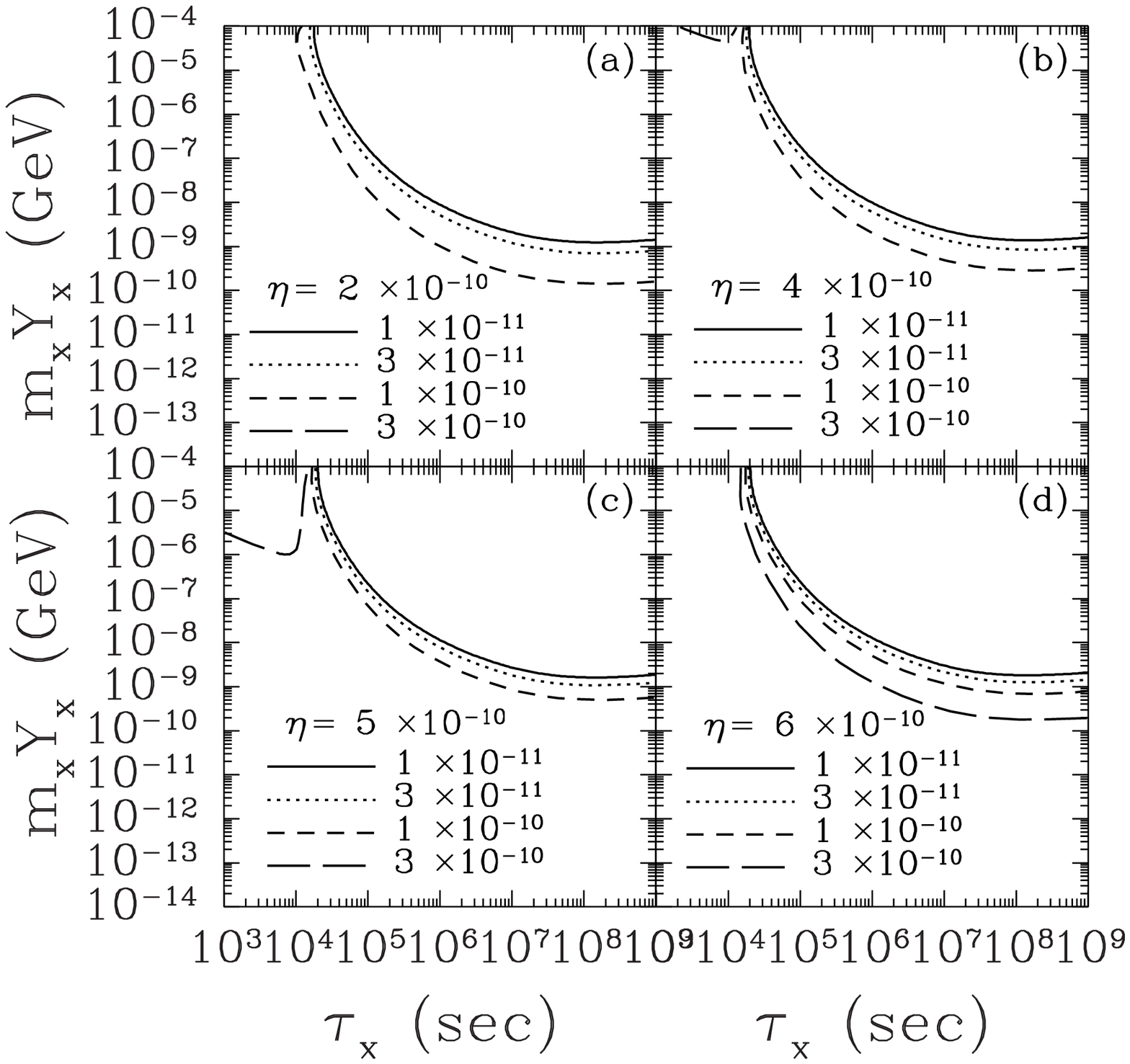}
\vspace{1cm}
\caption{
The abundance of $^7$Li/H, for the same
          theory parameters as in Fig.~5.
}
\label{fig:bbnx_7li}
\end{center}
\end{figure}
%

\newpage

%
\begin{figure}[hp]
\begin{center}
\vspace{2cm}
     \epsfxsize=1.0\textwidth\epsfbox{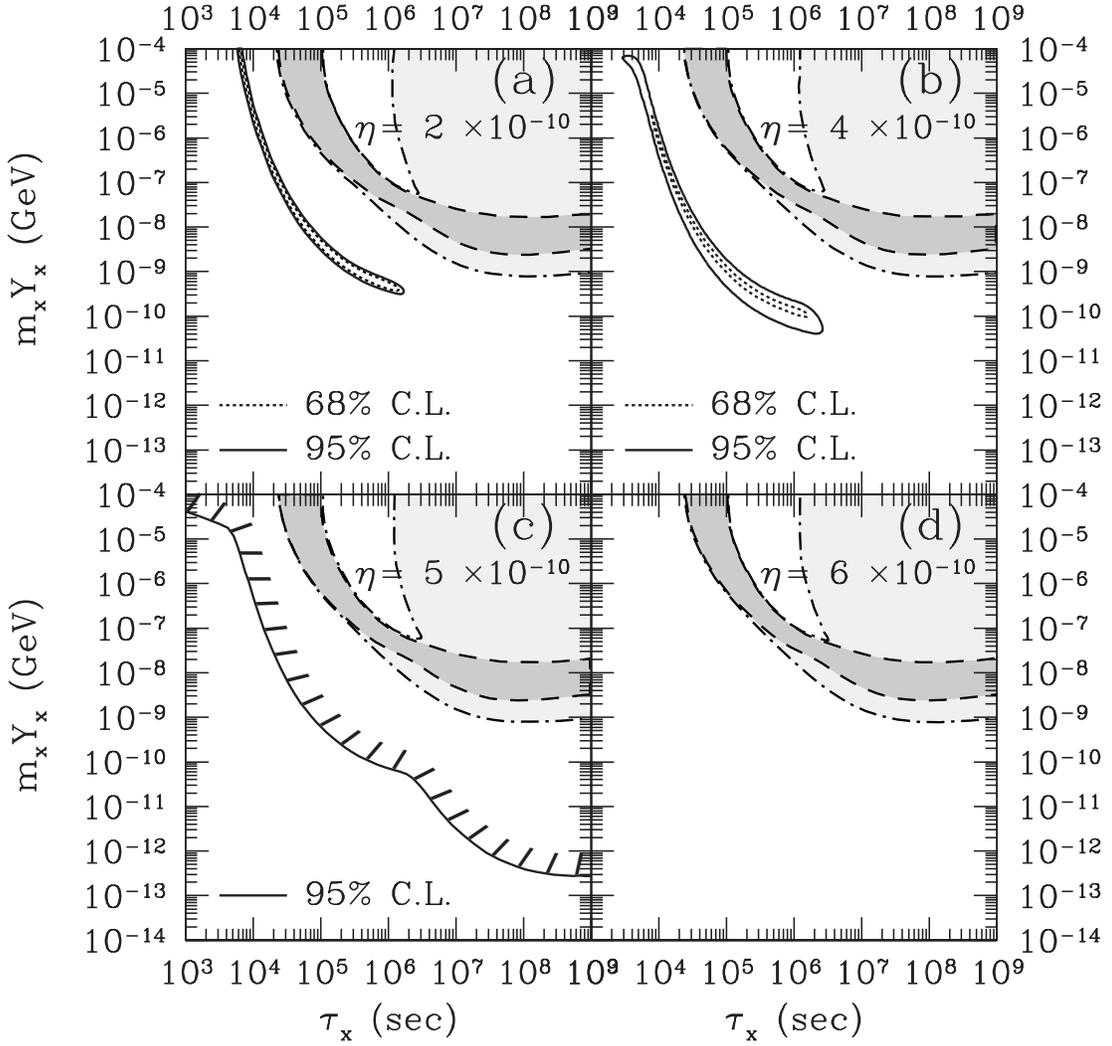}
\vspace{1cm}
\caption{
C.L. in the $m_X Y_X$ vs. $\tau_X$ plane, for low value of $Y$. We
take (a) $\eta=2\times 10^{-10}$, (b) $\eta=4\times 10^{-10}$, (c)
$\eta=5\times 10^{-10}$, and (d) $\eta=6\times 10^{-10}$. The shaded
regions are $y_6/y_7 \ge 0.5$, and the darker shaded regions are
$y_6/y_7 \ge 1.3$.
}
\label{fig:xchi2_ll}
\end{center}
\end{figure}
%

\newpage

%
\begin{figure}[hp]
\begin{center}
\vspace{2cm}
     \epsfxsize=1.0\textwidth\epsfbox{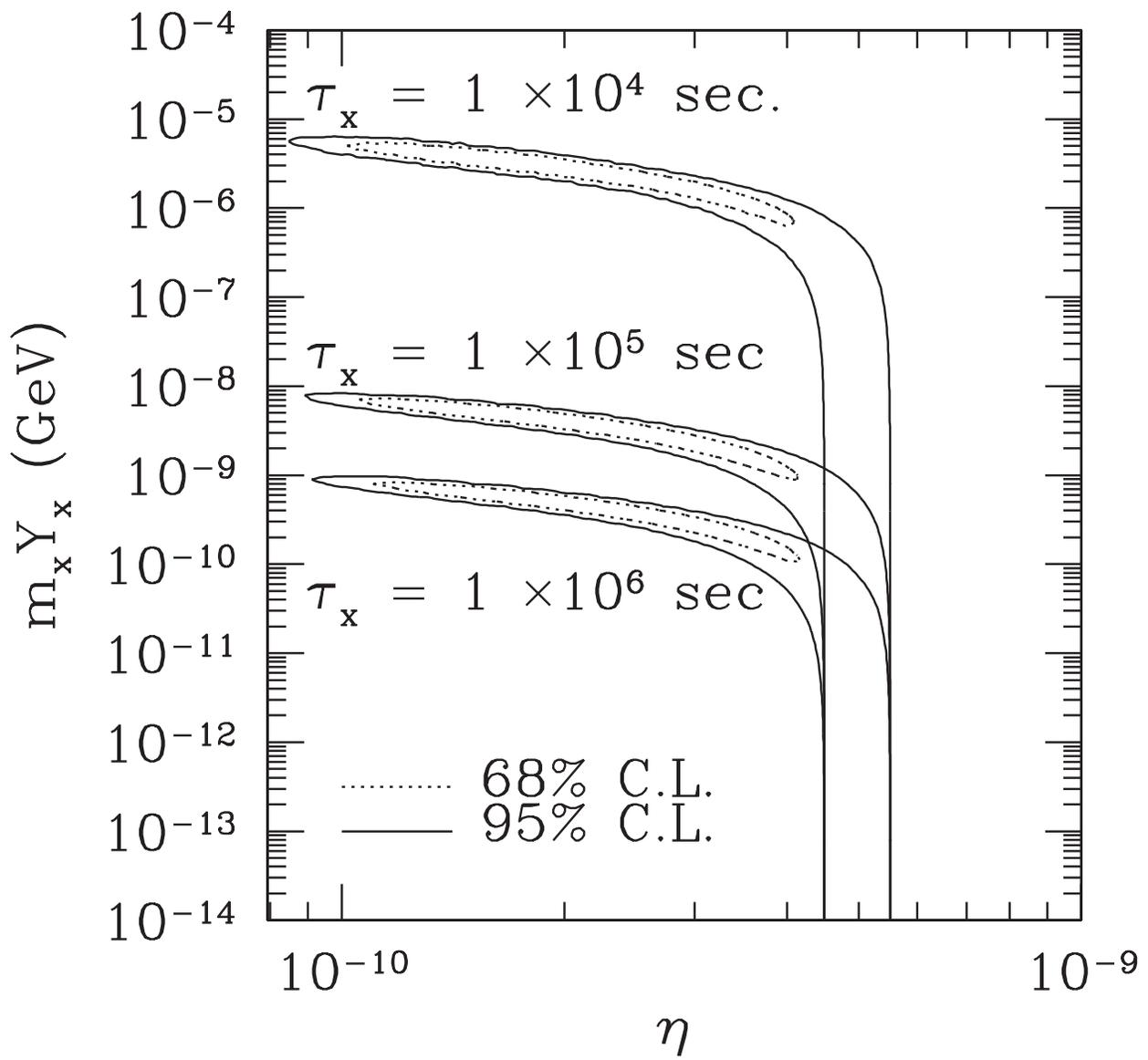}
\vspace{1cm}
\caption{
C.L. in the $\eta$ vs. $m_X Y_X$ plane for various values of
$\tau_X$, for low value of $Y$.
}
\label{fig:xtau_ll}
\end{center}
\end{figure}
%

\newpage

%
\begin{figure}[hp]
\begin{center}
\vspace{2cm}
     \epsfxsize=1.0\textwidth\epsfbox{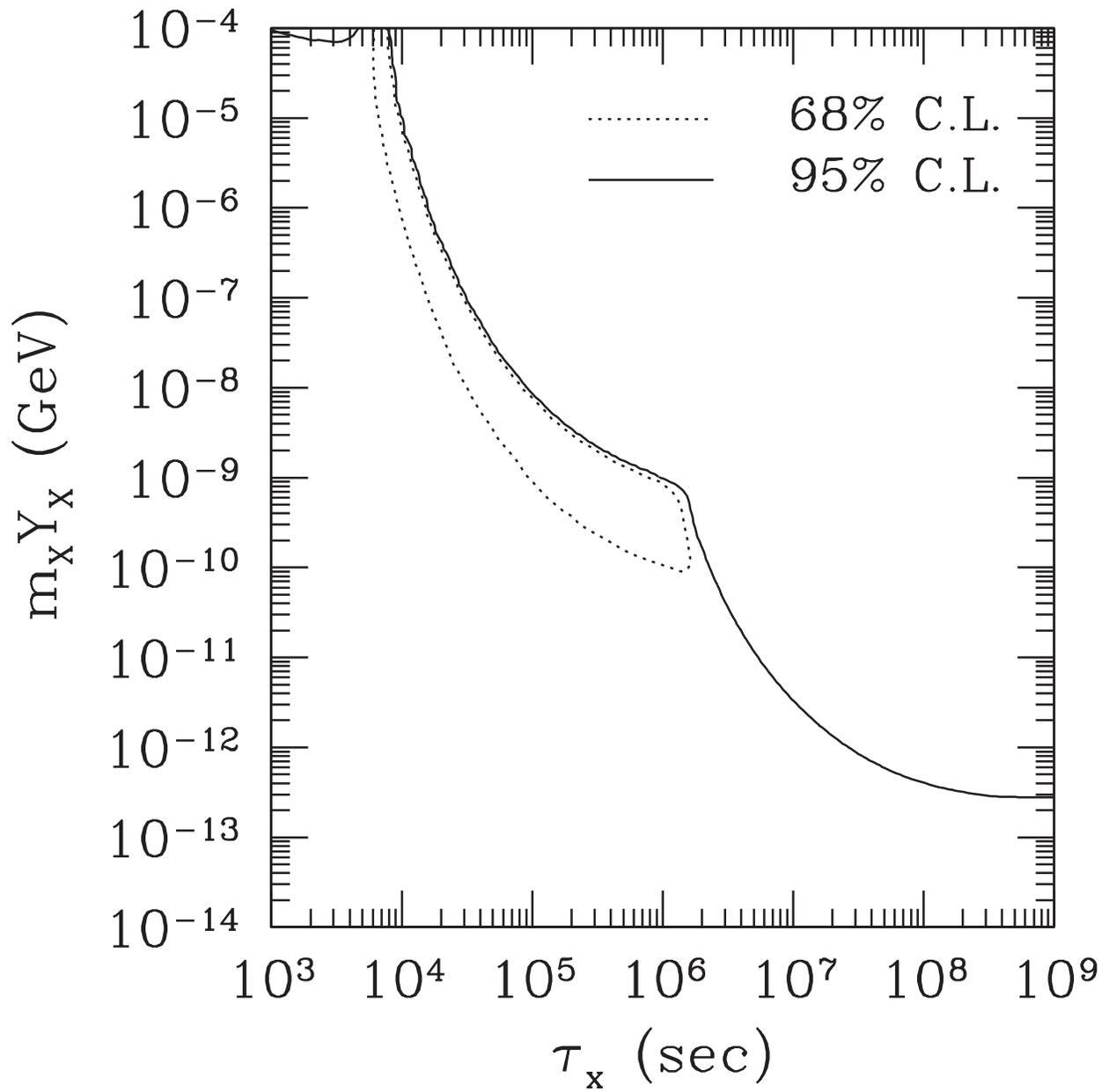}
\vspace{1cm}
\caption{
Contours of C.L. projected on $\eta$ axis for low value of $Y$.
}
\label{fig:xtot_ll}
\end{center}
\end{figure}
%

\newpage

%
\begin{figure}[hp]
\begin{center}
\vspace{2cm}
     \epsfxsize=1.0\textwidth\epsfbox{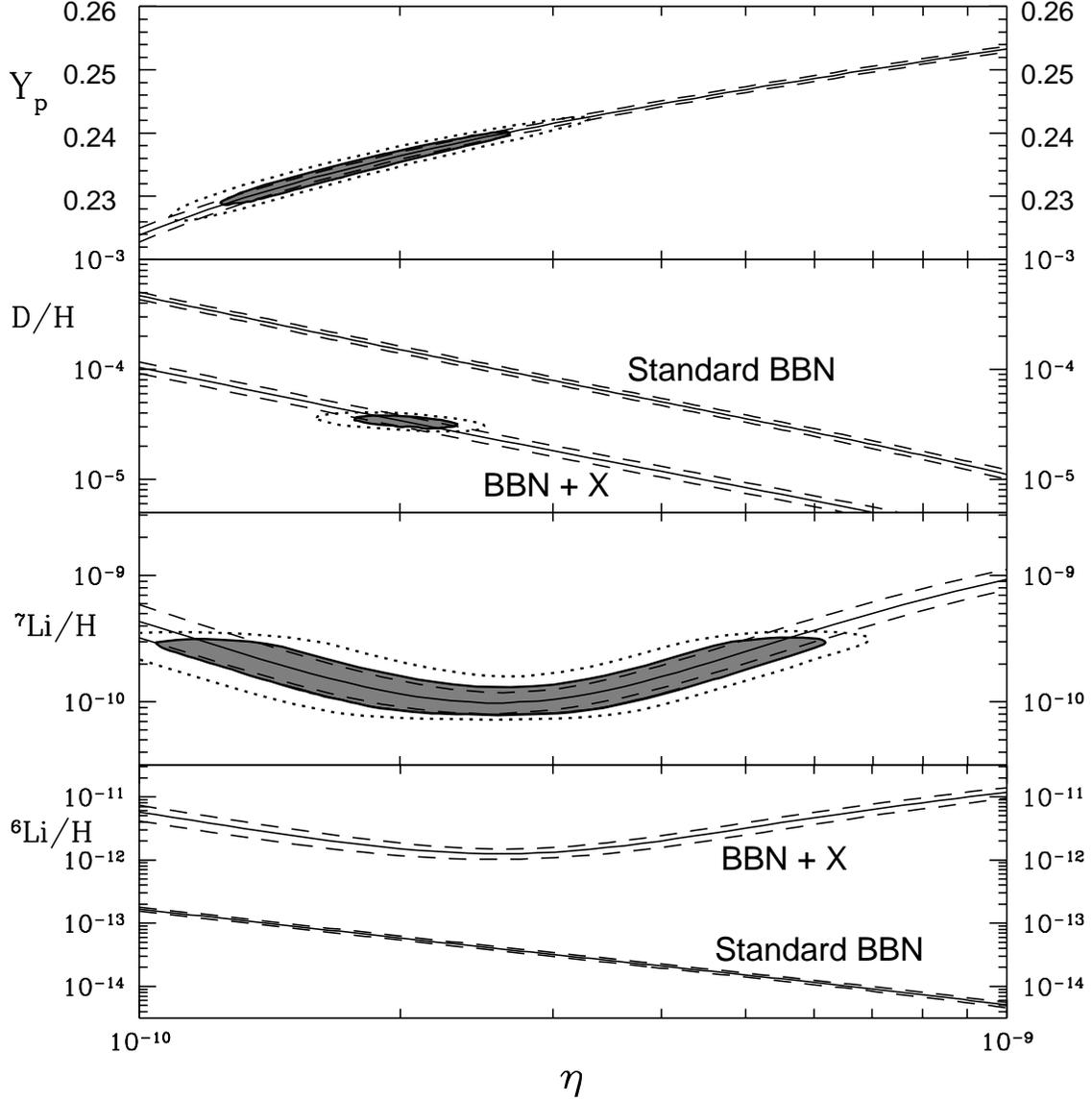}
\vspace{1cm}
\caption{
Predicted light-element abundances $^4$He, D, $^7$Li and $^6$Li at
$\tau_X = 10^6$ sec and $m_X Y_X = 5 \times 10^{-10}$ GeV. The
contours which are favored by observation are plotted, adopting the
low $^4$He and low D values. The dotted line denotes the 95\% C.L. and
the shaded region  denotes the 68\% C.L.. The predicted $^6$Li
abundance is two orders of magnitude larger   than the case of SBBN.
}
\label{fig:bbnxll}
\end{center}
\end{figure}
%

\newpage

%
\begin{figure}[hp]
\begin{center}
\vspace{2cm}
     \epsfxsize=1.0\textwidth\epsfbox{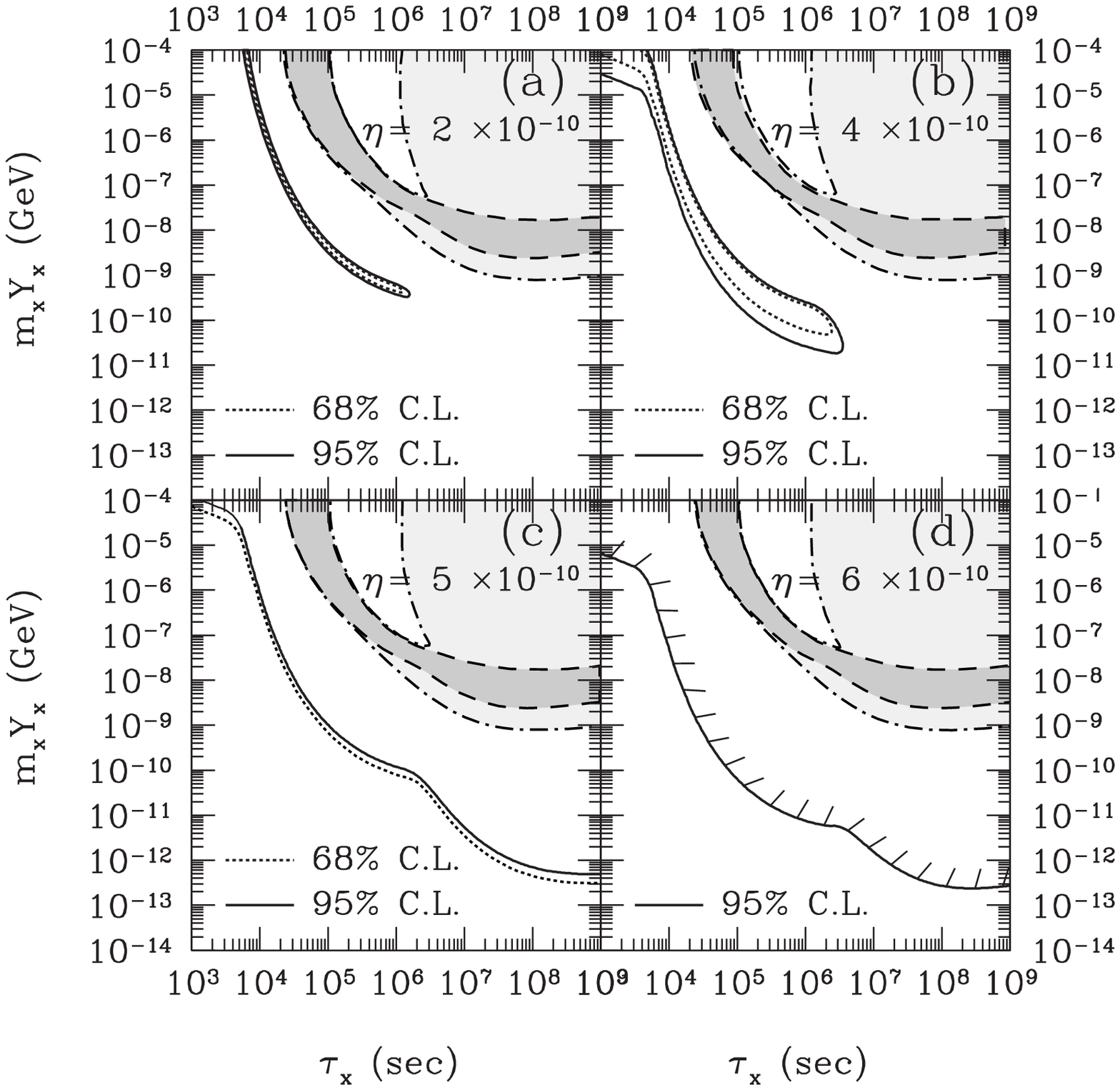}
\vspace{1cm}
\caption{
Same as Fig.~10, except for high value of $Y$.
}
\label{fig:xchi2_hh}
\end{center}
\end{figure}
%

\newpage

%
\begin{figure}[hp]
\begin{center}
\vspace{2cm}
     \epsfxsize=1.0\textwidth\epsfbox{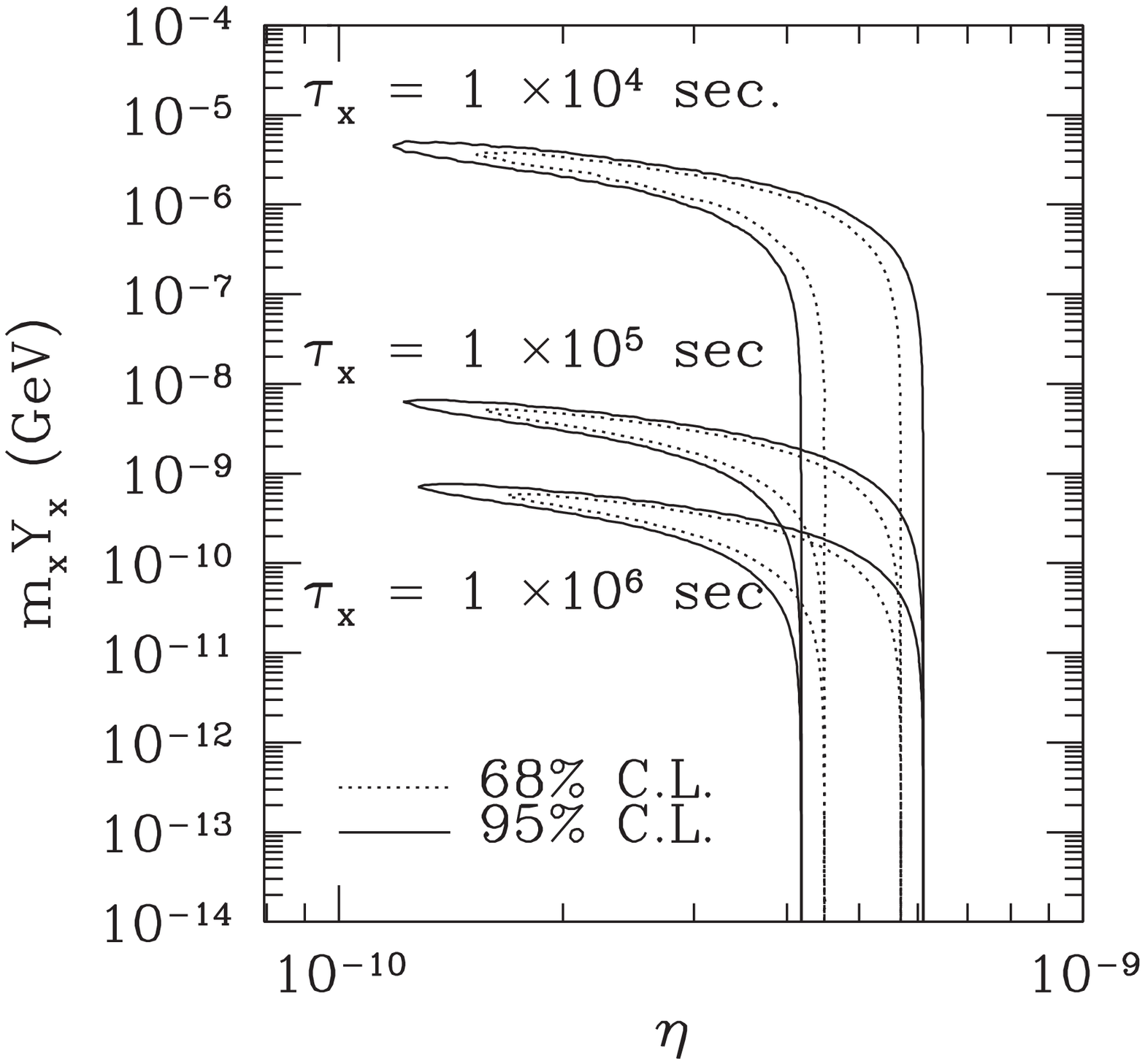}
\vspace{1cm}
\caption{
Same as Fig.~11, except for high value of $Y$.
}
\label{fig:xtau_hh}
\end{center}
\end{figure}
%

\newpage

%
\begin{figure}[hp]
\begin{center}
\vspace{2cm}
    \epsfxsize=1.0\textwidth\epsfbox{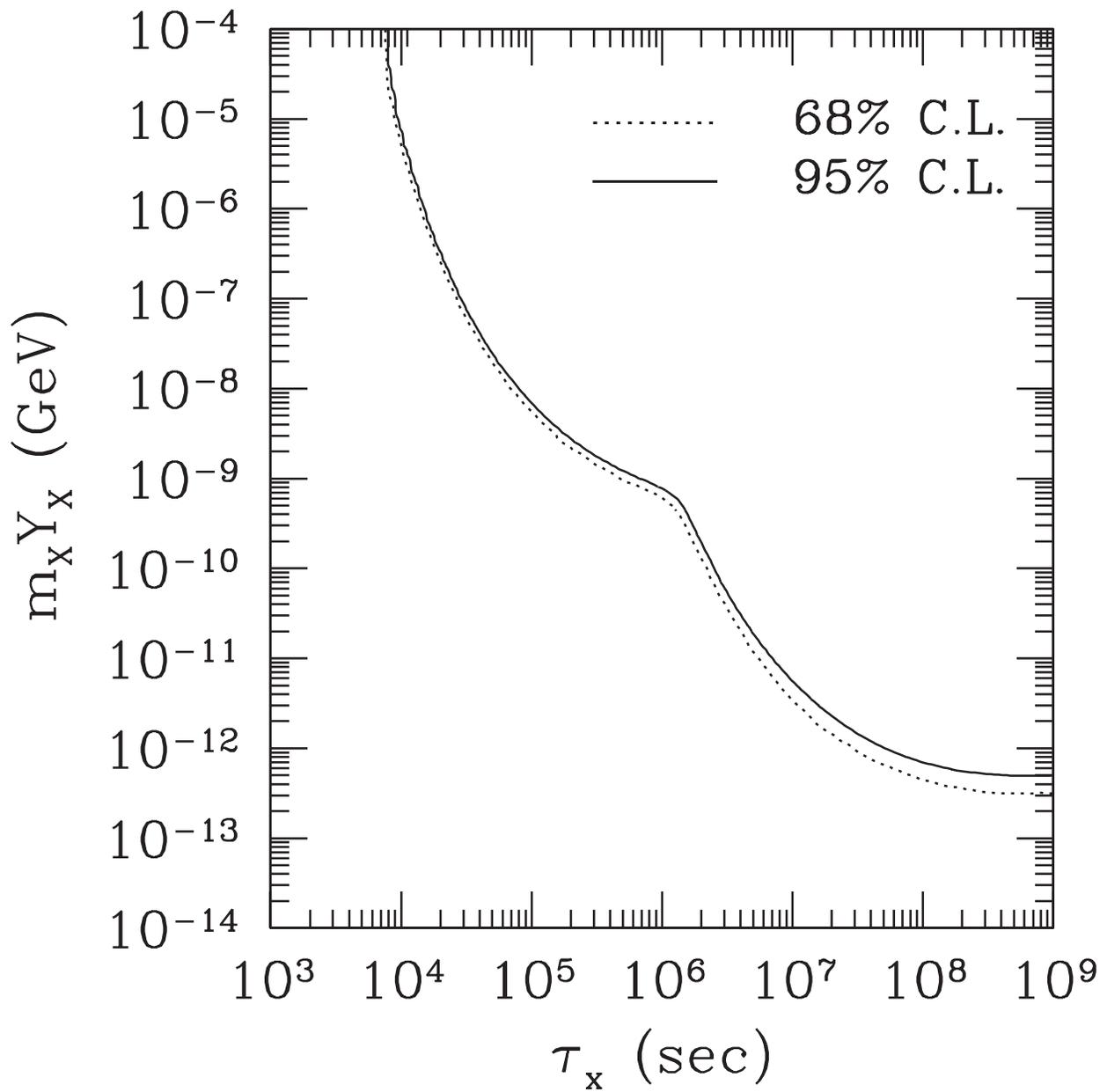}
\vspace{1cm}
\caption{
Same as Fig.~12, except for high value of $Y$.  The region above the
solid line is excluded at the 95\% C.L., while the region above the
dotted line is excluded at the 95\% C.L.
}
\label{fig:xtot_hh}
\end{center}
\end{figure}
%

\newpage

%
\begin{figure}[hp]
\begin{center}
\vspace{2cm}
    \epsfxsize=1.0\textwidth\epsfbox{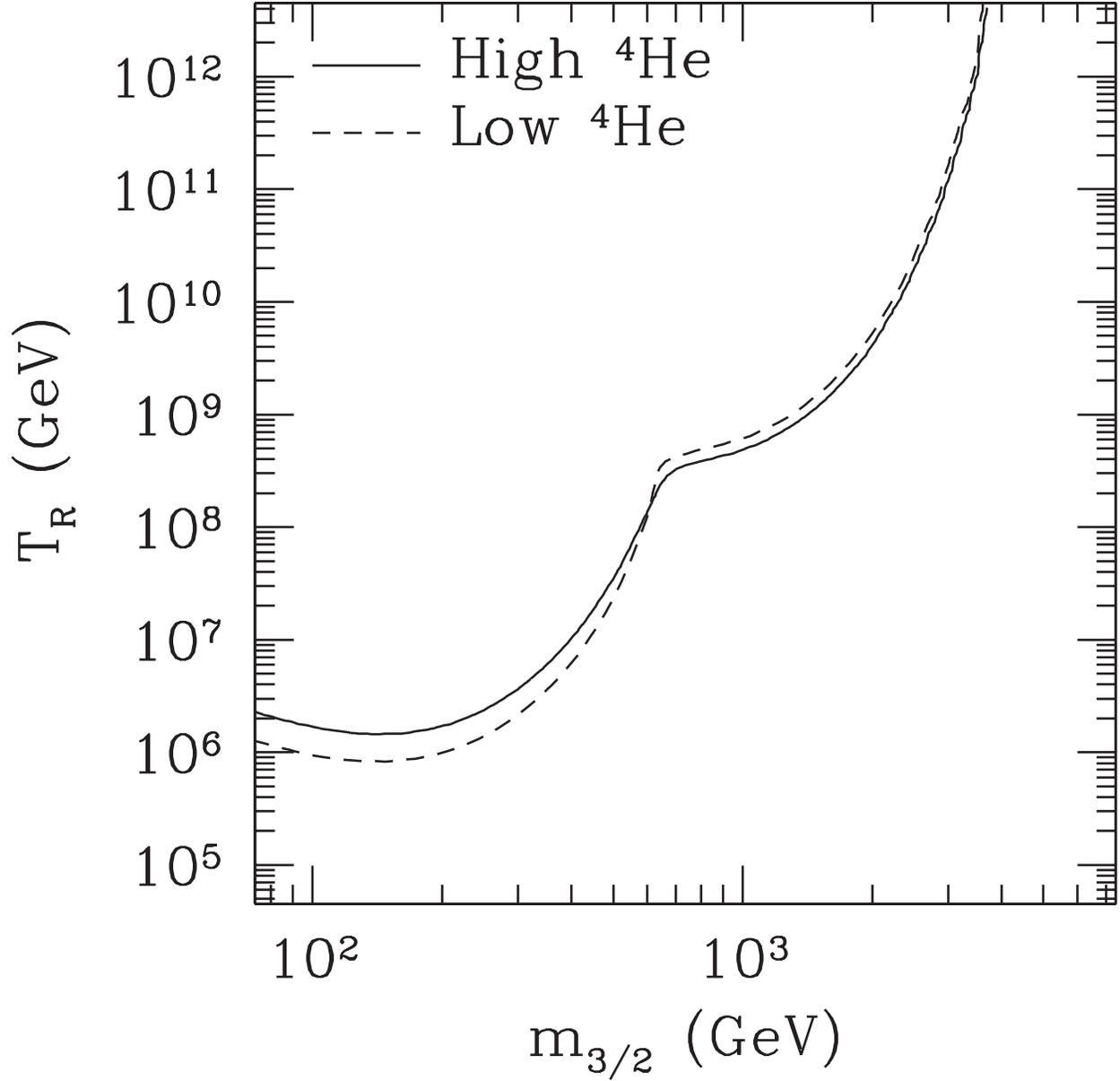}
\vspace{1cm}
\caption{
Contours of 95\% C.L., yielding an upper bound on the reheating
temperature, as a function of the gravitino mass. These are the
results of the two-photon emission by the decay of a gravitino. If we
consider the one-photon emission such as $\psi_{\mu} \to
\tilde{\gamma} + \gamma$, the upper bounds become milder by a factor
of two.
}
\label{fig:massTr}
\end{center}
\end{figure}
%

\newpage

%
\begin{figure}[hp]
\begin{center}
\vspace{3cm}
    \epsfxsize=1.0\textwidth\epsfbox{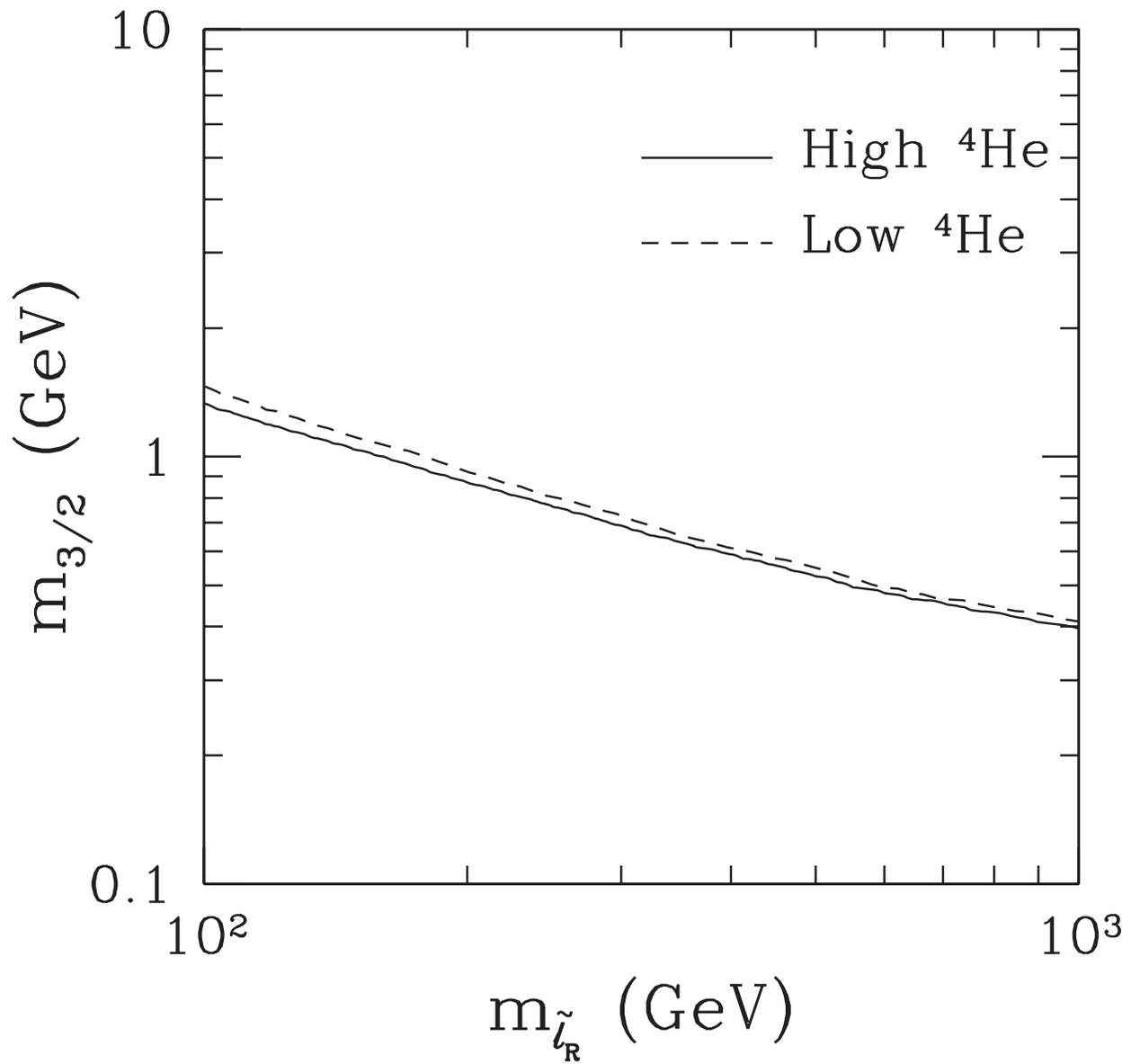}
\vspace{1cm}
\caption{
Contours of 95\% C.L., yielding an upper bound
on the gravitino mass, as a function of the right-handed slepton mass.
}
\label{fig:m32_mslR}
\end{center}
\end{figure}
%

\end{document}